\documentclass[fleqn,usenatbib]{mnras}

\usepackage{newtxtext,newtxmath,ulem}

\usepackage[T1]{fontenc}
\usepackage{ae,aecompl}
\usepackage{color}

\usepackage{graphicx}	
\usepackage{amsmath}	
\usepackage{lineno}

\usepackage{eso-pic}

\AddToShipoutPictureBG*{%
  \AtPageUpperLeft{%
    \hspace{0.75\paperwidth}%
    \raisebox{-3.5\baselineskip}{%
      \makebox[0pt][l]{\textnormal{DES-2022-0735}}
}}}%

\AddToShipoutPictureBG*{%
  \AtPageUpperLeft{%
    \hspace{0.75\paperwidth}%
    \raisebox{-4.5\baselineskip}{%
      \makebox[0pt][l]{\textnormal{FERMILAB-PUB-23-548-PPD}}
}}}%

\title[X-ray follow-up of DES Y3 redMaPPer]{Dark Energy Survey Year 3 Results: Mis-centering calibration and X-ray-richness scaling relations in redMaPPer clusters}

\author[DES Collaboration]{
\parbox{\textwidth}{
    \Large
P.~Kelly,$^{1,2}$ 
J.~Jobel,$^{1}$
O.~Eiger,$^{1,3,4}$
A.~Abd,$^{1,5}$
T.~E.~Jeltema,$^{1,3}$ 
P.~Giles,$^{6}$
D.~L.~Hollowood,$^{1,3}$
R.~D.~Wilkinson,$^{6}$
D.~J.~Turner,$^{6}$
S.~Bhargava,$^{6,7}$
S.~Everett,$^{8}$
A.~Farahi,$^{9}$
A.~K.~Romer,$^{6}$
E.~S.~Rykoff,$^{10,11}$
F.~Wang,$^{12,13}$
S.~Bocquet,$^{14}$
D.~Cross,$^{15}$
R.~Faridjoo,$^{16}$
J.~Franco,$^{1}$
G.~Gardner,$^{1}$
M.~Kwiecien,$^{1,3}$
D.~Laubner,$^{1,17}$
A.~McDaniel,$^{18}$
J.~H.~O'Donnell,$^{1,3}$
L.~Sanchez,$^{19}$
E.~Schmidt,$^{1}$
S.~Sripada,$^{20}$
A.~Swart,$^{21}$
E.~Upsdell,$^{6}$
A.~Webber,$^{18}$
M.~Aguena,$^{22}$
S.~Allam,$^{23}$
O.~Alves,$^{24}$
D.~Bacon,$^{25}$
D.~Brooks,$^{26}$
D.~L.~Burke,$^{10,11}$
A.~Carnero~Rosell,$^{27,22,28}$
J.~Carretero,$^{29}$
C.~A.~Collins,$^{30}$
M.~Costanzi,$^{31,32,33}$
L.~N.~da Costa,$^{22}$
M.~E.~S.~Pereira,$^{34}$
T.~M.~Davis,$^{35}$
P.~Doel,$^{26}$
I.~Ferrero,$^{36}$
J.~Frieman,$^{23,37}$
J.~Garc\'ia-Bellido,$^{38}$
G.~Giannini,$^{29}$
D.~Gruen,$^{14}$
R.~A.~Gruendl,$^{39,40}$
M.~Hilton,$^{41}$
S.~R.~Hinton,$^{35}$
K.~Honscheid,$^{42,43}$
D.~J.~James,$^{44}$
K.~Kuehn,$^{45,46}$
R.~G.~Mann,$^{47}$
J.~L.~Marshall,$^{48}$
J. Mena-Fern{\'a}ndez,$^{49}$
C.~J.~Miller,$^{50}$
R.~Miquel,$^{51,29}$
J.~Myles,$^{52,10,11}$
A.~Palmese,$^{53}$
A.~Pieres,$^{22,54}$
A.~A.~Plazas~Malag\'on,$^{10,11}$
P.~J.~Rooney,$^{6}$
M.~Sahlen,$^{55}$
E.~Sanchez,$^{49}$
D.~Sanchez Cid,$^{49}$
M.~Schubnell,$^{24}$
I.~Sevilla-Noarbe,$^{49}$
M.~Smith,$^{56}$
J.~P.~Stott,$^{57}$
E.~Suchyta,$^{57}$
M.~E.~C.~Swanson,$^{39}$
G.~Tarle,$^{24}$
C.~To,$^{42}$
P.~T.~P.~Viana,$^{58,59}$
N.~Weaverdyck,$^{24,60}$
P.~Wiseman$^{56}$
\begin{center} (DES Collaboration) \end{center}
}
}

\date{Accepted XXX. Received YYY; in original form ZZZ}

\pubyear{\today}

\begin{document}
\label{firstpage}
\pagerange{\pageref{firstpage}--\pageref{lastpage}}
\maketitle

\begin{abstract}
We use Dark Energy Survey Year 3 (DES Y3) clusters with archival X-ray data from XMM-Newton and Chandra to assess the centering performance of the redMaPPer cluster finder and to measure key richness observable scaling relations.  In terms of centering, we find that 10-20\% of redMaPPer clusters are miscentered with no significant difference in bins of low versus high richness ($20<\lambda<40$ and $\lambda>40$) or redshift ($0.2<z<0.4$ and $0.4 <z < 0.65$).  We also investigate the richness bias induced by miscentering. The dominant reasons for miscentering include masked or missing data and the presence of other bright galaxies in the cluster; for half of the miscentered clusters the correct central was one of the other possible centrals identified by redMaPPer, while for $\sim 40$\% of miscentered clusters the correct central is not a redMaPPer member with most of these cases due to masking.  In addition, we fit the scaling relations between X-ray temperature and richness and between X-ray luminosity and richness. We find a T$_X$-$\lambda$ scatter of $0.21 \pm 0.01$.  While the scatter in T$_X$-$\lambda$ is consistent in bins of redshift, we do find modestly different slopes with high-redshift clusters displaying a somewhat shallower relation.  Splitting based on richness, we find a marginally larger scatter for our lowest richness bin, $20 < \lambda < 40$. The X-ray properties of detected, serendipitous clusters are generally consistent with those for targeted clusters, but the depth of the X-ray data for undetected clusters is insufficient to judge whether they are X-ray underluminous in all but one case.

\end{abstract}




\section{Introduction}

The formation and evolution of galaxy clusters depends sensitively on the expansion history and matter density of the universe, and the growth of the cluster mass function with cosmic time can be used to constrain the dark energy equation of state \citep[e.g.][]{Allen11,Weinberg13,Huterer18}. With the advent of wide-field surveys such as the Dark Energy Survey (DES) and the Legacy Survey of Space and Time (LSST), optical cluster surveys can provide strong constraints on cosmology \citep[e.g.][]{Weinberg13, deskp}. The very large cluster samples in these surveys enable excellent statistical constraints, making control and calibration of systematics in cluster selection and characterization of the utmost importance. 

The cluster finder employed by the DES for cosmological analyses, redMaPPer \citep{redmapperI}, is a red sequence based algorithm for identifying clusters. RedMaPPer has been shown to have excellent redshift performance and its richness estimator, $\lambda$, has low intrinsic scatter with cluster mass \citep{Rozo09}.  The DES Y1 (year 1) cluster cosmology sample included a total of 7066 clusters in the volume-limited, $\lambda>20$ catalog.  The DES Y3 (years 1-3) $\lambda>20$, volume-limited cluster catalog forming the basis of this work in comparison contains 21,092 clusters. \cite{deskp} showed that the combination of cluster number counts and stacked cluster weak lensing in DES has the statistical potential to equal or exceed the constraining power of the DES 3x2pt combined analysis. However, even for DES Y1 the cluster analysis is systematics limited, and more specifically, these results imply the presence of unmodelled systematics for low-richness clusters.

Aspects of redMaPPer selection like miscentering and projection effects as well as richness scatter can be probed using multiwavelength follow up and spectroscopy.  Previous work has derived estimates for both the richness-mass scatter and miscentering fractions using Sunyaev-Zeldovich effect (SZ) \citep{saro15,bleem20}, X-ray \citep{redmapperSV,farahi19,matcha,kirby19,Zhang19,Giles} and weak-lensing (WL) \citep{mantz16,mulroy19} observations of redMaPPer clusters in SDSS, DES Science Verification (SV), and DES Y1.  Where available, spectroscopy has been used to probe projection effects and the velocity dispersion - richness relation \citep{farahi16,myles21,wetzell21}.  \cite{myles21} measure the fraction of observed richness contributed by non-cluster galaxies seen in projection for SDSS redMaPPer clusters; they find projection fractions generally consistent with the current DES models \citep*{Costanzi19a} but with a strong trend of increasing projection for decreasing richness.
In addition, SZ observations favor a different slope to the mass-richness relation than found with stacked weak lensing and hint at scatter or contamination that grows at low richness \citep{bleem20,costanzi21,grandis20}.

While the work of \cite{farahi19}, \cite{Zhang19}, and \cite{matcha} provides valuable X-ray follow-up and calibration for the DES cluster sample, previous studies are limited by small sample size and particularly by small samples of low-richness clusters and ``serendipitous'' clusters (clusters which are not imaged as the aimpoint of an X-ray follow-up observation). Likewise, current SZ samples are limited to high-richness clusters ($\lambda>40$). 
In this paper, we present X-ray scaling relations and mis-centering distributions for the DES Y3 clusters which have archival \textit{Chandra} or \textit{XMM-Newton} observations, including richness and redshift trends. We also explore the reasons for miscentering and the miscentering induced richness bias.  After flag and redshift cuts, our sample includes 676 unique clusters with X-ray observations of which 243 have high signal-to-noise detections; 82 of these detected clusters have richnesses $\lambda<40$. 

In Section 2, we briefly summarize the DES data and cluster catalog.  Section 3 presents both the \textit{Chandra} and \textit{XMM} data reduction, analysis, and X-ray cluster selection criteria.  In Section 4, we present our results for both cluster centering and X-ray-richness scaling relations in addition to discussing aspects of cluster selection and centering informed by the X-ray data. Throughout the paper, we assume a flat $\Lambda$CDM cosmological model with matter density $\Omega_{M} = 0.3$, Hubble constant $H_{0} = 70 \text{km} \cdot \text{s}^{-1} \cdot \text{Mpc}^{-1}$, and E(z), the dimensionless Hubble parameter as a function of redshift, defined as $E(z) = \frac{H(z)}{H_{0}} = \sqrt{\Omega_{M}(1+z)^{3} + 1 - \Omega_{M} }$.

\section{DES Data and Cluster Selection}

\subsection{DES Y3 Data}
In this work we characterize the X-ray properties of clusters selected from the Dark Energy Survey three year data set (DES Y3). 
DES Y3 
includes data acquired between 2013 August 15 and 2016 February 12 using the Dark Energy
Camera \citep[DECam][]{DECam} mounted on the Blanco 4m telescope at the Cerro Tololo Interamerican observatory in Chile. The DES Y3 GOLD catalog includes 388 million objects imaged in 4946 square degrees of the sky in the g, r, i, and z broadband filters \citep{Y3Gold}. 
This is a significant increase compared to the 1786 square degrees of sky included in the Y1 GOLD catalog from the first year of data \citep{Y1Gold}. 

\subsection{RedMaPPer}
\label{sec:desrm}

The red-sequence Matched-filter Probabilistic Percolation, or redMaPPer, cluster finding algorithm \citep{redmapperI}, has proven to be a powerful tool for selecting clusters from optical photometric survey data based on the cluster red sequence and has been applied to SDSS, DES Science Verification, and DES Y1 data \citep*{redmapperI, redmapperSV, mcclintock19}.  The redMaPPer algorithm iteratively self-trains the red sequence model using available spectroscopic redshifts, and furthermore iteratively calculates photometric redshifts for each cluster found. Starting from a potential central cluster galaxy, potential cluster galaxies are given a membership probability using  a matched-filter algorithm with filters for spatial position, color, and magnitude. This process is iterative with final membership probabilities determined relative to the most likely central galaxy. Up to five potential central galaxies are identified and given probabilities of being the cluster center.  The richness ($\lambda$) of each redMaPPer cluster is calculated as the sum of membership probabilities over all galaxies within a scale radius $R_{\lambda}$, where $R_{\lambda}=1.0h^{-1}{\rm Mpc}(\lambda / 100)^{0.2}$.

The cluster catalog used in this work was generated by redMaPPer v6.4.22+2 run on the DES Y3 data (Y3A2 Gold 2.2.1). The analysis in this paper was restricted to clusters in the volume-limited, $\lambda>20$ catalog containing 21,092 clusters.  We further limit the catalog to clusters in the redshift range $0.2<z<0.65$ adopted for DES cluster cosmology studies.

\section{X-ray analysis}
Starting from the DES Y3 redMaPPer cluster catalog, we utilize available, archival \textit{Chandra} and \textit{XMM-Newton} data at redMaPPer cluster positions to determine cluster X-ray properties including temperature, luminosity, X-ray centers, and luminosity upper limits for undetected clusters.  The X-ray data reduction and analysis are presented below for \textit{Chandra} and \textit{XMM} data in Sections 3.1 and 3.2, respectively.

Table~\ref{tab:data_summary} provides a summary of the \textit{Chandra} and \textit{XMM} samples used in this work, as outlined in Sections~\ref{sec:matcha} and~\ref{sec:xcs}.  Sample sizes for sub-samples studied are noted in the corresponding tables in Section 4.

\subsection{Chandra Sample and Analysis} \label{Chandra}
\label{sec:matcha}

We analyze archival \textit{Chandra} data for redMaPPer clusters using the \textbf{MATCha} (\textbf{M}ass \textbf{A}nalysis \textbf{T}ool for \textbf{Cha}ndra) pipeline as introduced in \citet{matcha}. Given a cluster catalog containing a set of equatorial coordinates (RA, Dec) and a redshift ($z$), MATCha automatically downloads any \textit{Chandra} data which includes these coordinates. It then attempts to find X-ray temperatures and luminosities ($T_X$, $L_X$) as well as cluster centroids and X-ray peaks. MATCha performs this analysis by running a series of CIAO (version 4.10 and CALDB version 4.8.1) \citep{ciao} and HEASOFT (version 6.24) tools\footnote{https://heasarc.gsfc.nasa.gov/ftools/} \citep{heasoft2}. In this section, we outline the data preparation and analysis in the MATCha algorithm and post-processing steps.

\subsubsection{MATCha Analysis}

MATCha first queries the \textit{Chandra} database for data overlapping the  \textit{redMaPPer} cluster positions using the CIAO tool \textit{find\_chandra\_obsid}. MATCha then downloads each set of \textit{Chandra} observations and reduces them using the CIAO tool \textit{chandra\_repro}. We narrow the energy range to 0.3-7.9 keV and remove particle background flares using the CIAO tool \textit{deflare} before creating images and exposure maps for each observation. MATCha identifies point sources using the CIAO tool \textit{wavdetect} and removes them from the data. 

MATCha then attempts to measure $T_X$, $L_X$, centroids and X-ray peaks. MATCha iteratively attempts to find a centroid within a 500 kpc region with the initial center at the redMaPPer position and with subsequent iterations centered at the most recent centroid. 
If no stable center is found within 20 iterations, MATCha marks the cluster as ``undetected'' and outputs an upper limit on $L_X$ using the redMaPPer position. Clusters whose signal-to-noise ratio are less than 5 within the final 500 kpc region are also marked as ``undetected.'' Otherwise, the cluster is marked as ``detected'', and we extract a spectrum within the 500 kpc radius using the CIAO tool \textit{specextract}. 

For detected clusters, we first fit $L_X$ and $T_X$ within the 500 kpc aperture using \textit{XSPEC} \citep{xspec} assuming a column density of galactic neutral hydrogen from the \textit{HEASOFT} tool \textit{nH}  (this is a weighted average of the hydrogen densities found in \cite{Kalberla} and \cite{Dickey}). Cluster spectra are fit using \textit{XSPEC}'s wabs*mekal model. The abundance is fixed to $0.3Z_{\sun}$, a value typical for X-ray clusters \citep{kravtsov12}, using the solar abundances from \cite{AndersGrevesse}. If the fit is successful, this process is repeated for additional aperture sizes, and centroids, $T_X$, and $L_X$ are found for $r_{2500}$, $r_{500}$, and core-cropped $r_{500}$ apertures (core size of $0.15 r_{500}$). These regions are also found iteratively, with initial guesses for $r_{2500}$ calculated from the 500 kpc $T_X$ and initial guesses for $r_{500}$ calculated from the $r_{2500}$ $T_X$. Here $r_\delta$ refers to the radius at which the mean mass density is $\delta$ times the critical density, and $r_\delta$ is estimated using the temperature-radius relations in \cite{arnaud05}.

For clusters which are detected but for which the temperature fit failed, MATCha will still estimate the luminosity.  In this case, the luminosity within a 500 kpc aperture is determined at a fixed temperature with iteration on the assumed $T_X$ using an $L_X-T_X$ relation to guess the temperature based on the measured luminosity.  Initially, the luminosity is determined for a starting $T_X$ of 3 keV; the temperature guess is updated based on the $L_X-T_X$ relation and the luminosity determined for the new temperature.  This process is repeated until $L_X$ is unchanged within the uncertainties.  In this work, we use an $L_X-T_X$ relation based on previous fits to Chandra observations of SDSS redMaPPer clusters \citep{matcha}.  500 kpc is roughly equivalent to $r_{2500}$ for our clusters \citep{matcha}.  For undetected clusters, we estimate a $3\sigma$ upper limit on $L_X$ within a 500 kpc aperture with an assumed $T_X = 3$ keV.

MATCha additionally determines X-ray peaks for detected clusters by determining the brightest pixel in smoothed, exposure-corrected, point source subtracted images. Images are smoothed using a $\sigma=50$ kpc Gaussian.

\subsubsection{Post processing}

Several problematic cases or failure modes of the automated analysis are identified in the post processing.  These include instances in which most of the cluster source or background regions are not in \textit{Chandra}'s field of view and clusters for which \textit{Chandra} was not in an imaging mode when all images were taken. Separate flags were used for proximity to chip edges affecting determination of the X-ray center, $r_{2500}$, $r_{500}$, or background properties. Clusters where
a second nearby cluster significantly contaminated the emission or background were additionally flagged. We also check that no detected clusters are actually bright nearby clusters rather than the intended \textit{redMaPPer} cluster by comparing the output cluster catalog to the NASA/IPAC Extragalactic Database\footnote{https://ned.ipac.caltech.edu/ The NASA/IPAC Extragalactic Database (NED) is funded by the National Aeronautics and Space Administration and operated by the California Institute of Technology.} as well as other redMaPPer clusters in the field. These clusters are flagged as ``masked" and cut from the sample.

Post-processing visual analysis is also required to check the locations of the X-ray peak positions. If the centers are incorrect due to point-source emission, the point source subtraction is adjusted and the center corrected. Mispercolations \citep[as detailed in][]{matcha} are identified visually and corrected by assigning the higher richness in redMaPPer to the more luminous X-ray cluster. The less luminous cluster is flagged as masked and removed from the sample. In practice, there was only one such case in the Y3 Chandra sample.  For a second rich cluster in the sample, the X-ray position was misidentified as a $\lambda < 20$ cluster; we simply treat this cluster as miscentered.

\subsubsection{Chandra Samples and Flag Cuts}
\label{sec:chandracuts}

There were 415 clusters from the volume-limited, $\lambda>20$ Y3 redMaPPer catalog within $0.2<z<0.65$ falling with archival Chandra data.  Of these, 186 clusters were detected.  We further restrict this sample based on signal-to-noise ratio (SNR) as this improves the centering accuracy and decreases the point source contamination. Specifically, we remove clusters with a SNR in a 500 kpc aperture less than 9.0, which was determined to be a good minimum for both the Chandra and XMM samples, cutting 46 clusters. In addition, for determination of the X-ray scaling relations we removed clusters flagged as masked by another cluster, bad (non-imaging) mode, lying too close to a chip edge for robust determination of the X-ray temperature or background, and clusters where another nearby cluster significantly contaminated the emission sampled in either the cluster or background regions.  In total an additional 27 clusters were removed by the flag cuts, leaving 113 clusters.  For 15 of these clusters, we were unable to fit an $r_{2500}$ temperature due to poor statistics in the spectrum, leaving a final sample of 98 clusters used to fit the $T_{X,r2500}-\lambda$ scaling relation. For the $r_{500}$ aperture, the same flags were applied with the only difference in cuts being 2 clusters where an $r_{500}$ temperature could not be fit, leaving 96 clusters to be used in the $T_X-\lambda$ fit for an $r_{500}$ region.

For cluster X-ray luminosity and the determination of the $L_X-\lambda$ relations, there are several different cases to consider.  We utilize the measured luminosities for 113 clusters with SNR $>9$ and meeting the same flag cuts above, of which 98 have a measured $T_X$ and 15 have luminosities determined using an estimated $T_X$ from iteration on the $L_X-T_X$ relation. For both undetected clusters and detected clusters with SNR $<9$, we utilize only upper limits on the luminosity and include these data as censored in fits of the $L_X-\lambda$ relation.  This category includes 183 undetected clusters and 17 clusters with $5 <$ SNR $<9$, giving a total of 313 clusters in the luminosity sample.


Less restrictive cuts were used for the centering analysis in Section \ref{sec:desy3cent} as here it was only necessary that the X-ray peak position could be robustly determined.  The same cuts for SNR, masking, and bad mode were used, but clusters near chip edges were only flagged if the cluster was close enough to the edge to affect peak determination, and clusters with neighboring clusters were only flagged if the nearby cluster was within the $r_{2500}$ region.  These cuts gave a total sample of 124 clusters for the centering analysis.

\subsection{XCS Sample and Analysis}
\label{sec:xcs}

Here we describe the construction of the DES Y3 cluster sample cross-matched with available \textit{XMM} data.  The \textit{XMM} data are made available from the \textit{XMM} cluster survey \citep[XCS,][]{romer01}. The aim
of XCS is to catalogue and analyse all X-ray clusters detected during the XMM mission. The XCS data used in this work comprises all publicly available \textit{XMM} observations as of April 2019\footnote{collected from http://nxsa.esac.esa.int/nxsa-web/}.  Much of the process outlined here is detailed in \cite{Giles}, with the main matching and analysis process briefly described below. 

\subsubsection{\textit{XMM} reduction, Image Generation and Source Detection}
\label{sec:xmmreduction}

The XCS reduction process is fully described in \citet[][hereafter LD11]{xcs11}, and we outline the process here.  The data were processed using XMM-SAS version 14.0.0, and events lists generated using the {\sc EPCHAIN} and {\sc EMCHAIN} tools.  Periods of high background levels and particle contamination were excluded using an iterative 3$\sigma$ clipping process performed on the light curves, and time bins falling outside this range excluded.

Single camera (i.e. PN, MOS1 and MOS2) images and exposure maps were then generated from the cleaned events files, spatially binned with a pixel size of 4.35$^{\prime\prime}$.  The images and exposure maps were extracted in the 0.5 -- 2.0 keV band, with individual camera images and exposure maps merged to create a single image per ObsID.  The MOS cameras were scaled to the PN by the use of energy conversion factors (ECFs) derived using {\sc xspec}.  The ECFs were calculated based upon an absorbed power-law model.    

Using the merged images and exposure maps, the XCS Automated Pipeline Algorithm ({\sc xapa}) was used for source detection.  Once again, full details of the analysis can be found in LD11.  {\sc xapa} uses a wavelet process based upon the {\sc wavdetect} package \citep{freeman2002}.  Proceeding source detection, detected sources are classified as either point-like or extended.  After removal of duplicated sources (i.e. the same sources detected at different epochs), an XCS master source list (MSL) is created.  The XCS MSL used in this work contains 338,417 X-ray sources, of which 36,710 are classed as extended.

\subsubsection{DESY3 cross-match with \textit{XMM} archive}
\label{sec:xmmmatch}

The DESY3 redMaPPer sample defined in Sect~\ref{sec:desrm} was matched to \textit{XMM} images with the requirement that the redMaPPer position falls within 13$^{\prime}$ of the aimpoint of an \textit{XMM} observation.  Next, \textit{XMM} images were removed from further analysis based upon conditions on the total cleaned exposure time.  \textit{XMM} images were removed if, within a 5 pixel radius (centered on the redMaPPer position), the mean exposure time is <3ks and the median exposure time is <1.5ks.  The median exposure was employed in order to exclude redMaPPer clusters whose X-ray position would be significantly affected by, e.g. chip gaps.  Finally, the same exposure cut was carried out at a position 0.8R$_{\lambda}$ away from the redMaPPer position (in the direction away from the centre of the \textit{XMM} observation).  This is done for two reasons; (i) to reduce the number of clusters near the edge of the field-of-view of XMM; and (ii) to encapsulate the mis-centering measured between the redMaPPer central galaxy and the X-ray peak position as found in \cite{Zhang19}, and further explored in Section~\ref{sec:desy3cent}.  Based on these requirements, there were 1052 DESY3 redMaPPer clusters that fell within the footprint of an \textit{XMM} image (this sample is denoted as the DESY3RM-XMM sample).

\subsubsection{Generation of the DESY3RM-XCS catalogue}
\label{sec:desy3-xcs}

We cross-matched the 1052 DESY3RM-XMM clusters (see Sect.~\ref{sec:xmmmatch}) with the XCS MSL.  At the position of the redMaPPer defined central position, we match to all {\sc xapa} defined extended sources within a co-moving distance of 2~$h^{-1}$~Mpc, calculated at the redshift of the redMaPPer cluster in question.  We made the assumption that the closest {\sc XAPA} extended source was associated with the DESY3RM cluster in question. These matches were visually inspected to confirm the association of the extended XCS source with the DESY3RM cluster.  Additionally, the visual inspection process was used to identify X-ray observations effected by e.g., periods of high background, these were subsequently removed from further analysis.

To perform the inspection process, we inspect both DES Y3 and \textit{XMM} images.  In order to assess whether the XCS extended source is a true match to the redMaPPer cluster, we highlight both the position redMaPPer galaxies associated with the cluster in question, and the galaxies of other nearby redMaPPer clusters to ensure the X-ray emission is not associated with an redMaPPer cluster in projection.  Confirmed matched redMaPPer clusters are retained, with the sample containing 325 clusters. The remaining 697 DESY3RM clusters have no associated extended X-ray source within 2 h$^{-1}$ Mpc.

As done the Chandra sample (see Section~\ref{sec:chandracuts}), we further restrict the XMM confirmed matched clusters by removing clusters with a SNR in a 500 kpc aperture less than 9.0, cutting 70 clusters.  Finally, only clusters from the volume-limited Y3 redMaPPer catalogue, within 0.2$<$z$<$0.65, are retained, resulting in a final sample of 161 clusters.

\subsubsection{XCS Spectral Analysis}
\label{sec:xcsspec}

All \textit{XMM} data were analysed using the XCS Post Processing Pipeline \citep[XCS3P, see LD11 and updates in][]{Giles}.  Cluster spectra were extracted and fit using {\sc xspec}.  The fits are performed in the 0.3-7.9 keV band with an absorbed {\tt APEC} model \citep{apec} using the $c$-statistic \citep{cash1979}.  The abundance was fixed at 0.3$Z_{\odot}$ and the redshift fixed at the value of the redMaPPer defined redshift (note that redshift uncertainties are not taken into account in the fit), leaving the {\tt APEC} temperature and normalisation free to vary.  The absorption due to the interstellar medium was taken into account using a multiplicative {\tt Tbabs} model, with the value of the absorption ($n_{H}$) taken from \cite{2016A&A...594A.116H} and frozen during the fitting process.  The {\tt APEC} temperature and normalisation were free to vary during the fitting process.  Temperature errors were estimated using the {\tt XSPEC} {\tt ERROR} command, and quoted within 1-$\sigma$. Finally, luminosities (and associated 1-$\sigma$ errors) were estimated from the best-fit spectra using the {\tt XSPEC} {\tt LUMIN} command.  Spectra for each individual XMM camera were extracted and we filtered out spectra that did not, individually, produce a fitted temperature (complete with $1\sigma$ upper and lower limit values) in the range 0.08 $<$ $T_{X}$ $<$ 20 keV \citep[see][Section 3.1.2]{Giles}.  This was performed during each step in the iteration process outlined below.

Spectra are extracted within $r_{2500}$ as done for the MATCha analysis with values estimated using the following equation,
\begin{align}
\label{equ:r2500}
E(z)_{r2500} &= B_{\delta}\left( \frac{T_{\rm X}}{5 {\rm keV}} \right)^{\beta},
\end{align}
where $B_{\delta}$=491~kpc and $\beta$=0.56, taken from \cite{arnaud05}.  The local background was taken into account using an annulus centered on the cluster with an inner and outer radius of 2$r_{2500}$ and 3$r_{2500}$ respectively.  An initial temperature was calculated within the XAPA defined source region and this is used to estimate $r_{2500}$ using Equation~\ref{equ:r2500}.  A new $T_{\rm X}$ value is defined within this $r_{2500}$, and this is in turn used to define a new $r_{2500}$ value. The process was repeated until $r_{2500}$ converges (the ratio of the new to old $r_{2500}$ defined to be $>$0.9 and $<1.1$).  We required the iteration process to iterate at least three times (irrespective of convergence), up to a maximum of ten iterations.  If convergence was not achieved after ten iterations, the process was stopped and no temperature obtained.  The same iteration procedure was followed to extract $r_{500}$ temperatures.  However, for the $r_{500}$ analysis, the coefficients in Equation~\ref{equ:r2500} were $B_{\delta}$=1104~kpc and $\beta$=0.57 and the local background used an annulus with an inner and outer radius of 1.05$r_{500}$ and 1.5$r_{500}$ respectively.  For a few clusters with successful $r_{2500}$ the $r_{500}$ fits did not converge, giving a slightly smaller sample for this aperture.

For clusters where the iteration process failed, we estimate a luminosity with a fixed temperature, using the process outlined in \citet[][Section 3.2]{Giles}.  Briefly, the same iteration process is used as above, with the temperature fixed during the fitting process.  Initially, a spectrum is extracted and a temperature of 3~keV was used in the spectral fit to estimate a luminosity.  This luminosity was then used to estimate a new temperature using the $L_{X}$-$T_{X}$ relation given in \citet[][see Table 3]{Giles}.  An updated $r_{500}$ is calculated and the process repeated until convergence (as above). 
 Lastly, for undetected clusters, we estimate a $3\sigma$ upper limit on $L_X$ within a 500 kpc aperture with an assumed $T_X = 3$ keV, using the {\sc eregionanalyse} tool.

\begin{table}
\begin{center}
\begin{tabular}{|l|c|c|c|c|}
\hline
Sample & $N_{\rm sam}$ & $z_{\rm med}$  & $\lambda_{\rm med}$ & $T_{\rm X, med}$ [keV]  \\ \hline
Chandra (centering) & 124  & 0.39 & 98  & 7.11  \\ 
Chandra ($T_{\rm X, r_{2500}}$) & 98 & 0.40 & 105 & 7.23 \\
XMM ($T_{\rm X, r_{2500}}$)  & 161 & 0.39 & 47 & 3.96 \\
\hline
\end{tabular}
\caption{Summary of the X-ray samples used in this work and their median redshift, richness, and X-ray temperatures.}
\label{tab:data_summary}
\end{center}
\end{table}


\section{redMaPPer Centering}
\label{sec:desy3cent}

RedMaPPer defines the cluster position to be the location of the redMaPPer-determined most likely central galaxy.  In most cases, this galaxy is the correct central, but in a non-negligible fraction of cases redMaPPer miscenters choosing the wrong galaxy as the central \citep{saro15,matcha,Zhang19,bleem20}.  Miscentering biases the stacked weak lensing cluster mass estimates as well as the measured richnesses, but given a model for miscentering, we can calibrate for these effects.  The important parameters here are the fraction of clusters which are miscentered and the distribution of miscentering distances. 

\subsection{Centering Methods}
\label{sec:desy3centmethods}
We use the X-ray information to probe miscentering in two ways: First, we measure the offsets between the cluster X-ray peak position and the nominal redMaPPer central galaxy, and model the offset distribution using a two component model, one for well-centered and one for miscentered clusters. Second, we use the X-ray contours to visually identify the correct central cluster galaxy and compare this to the redMaPPer choice.  As detailed in Sect.~\ref{sec:desy3centresults}, these methods agree well in terms of the fraction of miscentered clusters and their offset distribution.  The X-ray emitting ICM, comprising the bulk of the baryonic component, serves as a good proxy to the gravitational center of clusters as the ICM density traces the gravitational potential and outside of short-lived periods near major mergers the ICM is largely in hydrostatic equilibrium within this potential.

The X-ray peak to redMaPPer offsets are measured relative to the richness-dependent radius $R_{\lambda} = (\lambda/100)^{0.2} h^{-1}$ Mpc. Following \citet{Zhang19}, we model the X-ray peak to redMaPPer central galaxy offset distribution using a two component model of the form

\begin{equation}
\begin{split}
P(x | \rho, \sigma, \tau) &=\rho \times 
P_\mathrm{cen}(x | \sigma) + (1-\rho) \times P_\mathrm{mis}(x | \tau), \\
P_\mathrm{cen}(x | \sigma)& = \frac{1}{\sigma} \mathrm{exp}(- \frac{x}{\sigma}), \\
P_\mathrm{mis}(x | \tau) &= \frac{x}{\tau^2} \mathrm{exp}(- \frac{x}{\tau}).\\
\label{eq:offset}
\end{split}
\end{equation}

Here $\rho$ is the fraction of well-centered clusters; $P_\mathrm{cen}$ is a Gamma distribution of shape parameter 1 and scale $\sigma$ representing the offset distribution of well-centered clusters, $P_\mathrm{mis}$ is a Gamma distribution of shape parameter 2 and scale $\tau$ representing the offset distribution of miscentered clusters, and $x$ is the X-ray to redMaPPer position offset normalized by $R_{\lambda}$.  Some width of the distribution for well-centered clusters, quantified here by $\sigma$, is expected due to the finite resolution of both the X-ray and optical images and potential gas-galaxy offsets following mergers.

In addition to probing the X-ray to redMaPPer offset, we used the X-ray and DES images to identify, where possible, the correct central cluster galaxy.  This analysis revealed the typical reasons for miscentering and allowed us to fit the true central to redMaPPer central distribution for miscentered clusters.  Each cluster in the centering samples for both Chandra and XMM were examined visually using the X-ray contours, DES images, and redMaPPer membership information.  Typically each cluster was looked at by two people, and flagged clusters were additionally checked by TJ.  Clusters were flagged if redMaPPer clearly picked the wrong galaxy as the central or if the determination of the central was ambiguous (e.g. there were multiple bright galaxies that could be the central(s)). 

The redMaPPer algorithm outputs the positions and centering probabilities of the five most likely central galaxies with the nominal center taken as the position of the most likely central. For clusters where redMaPPer chose the wrong galaxy, we recorded the DES position of the correct central galaxy and additionally flagged cases where the correct central was one of the other 4 possible central galaxies identified by redMaPPer and cases where the correct central was not a candidate member of any redMaPPer cluster. ``Ambiguous" clusters identify those for which it is ambiguous whether the redMaPPer chosen central is correct; in general, there can be cases where the redMaPPer position is clearly wrong but the central galaxy identification is still ambiguous, but for clarity we only list these as miscentered.  The designation ``Miscentered" was reserved for cases where redMaPPer clearly chose a galaxy outside of the cluster core or in a clearly subdominant structure outside the main halo. 

We note that the methodology for associating X-ray emission to redMaPPer clusters may lead us to miss clusters with very large miscentering distances.  In DES Y3, we find one example, missed in both the Chandra and XMM samples, where Abell 209 is found as a low-richness redMaPPer cluster offset by 2.4 Mpc from the X-ray center due to gap in the DES data \citep[see Figure 13 of][]{wetzell21}.

\subsection{Centering Results}
\label{sec:desy3centresults}

\begin{figure*}
    \centering
    \includegraphics[width=8cm]{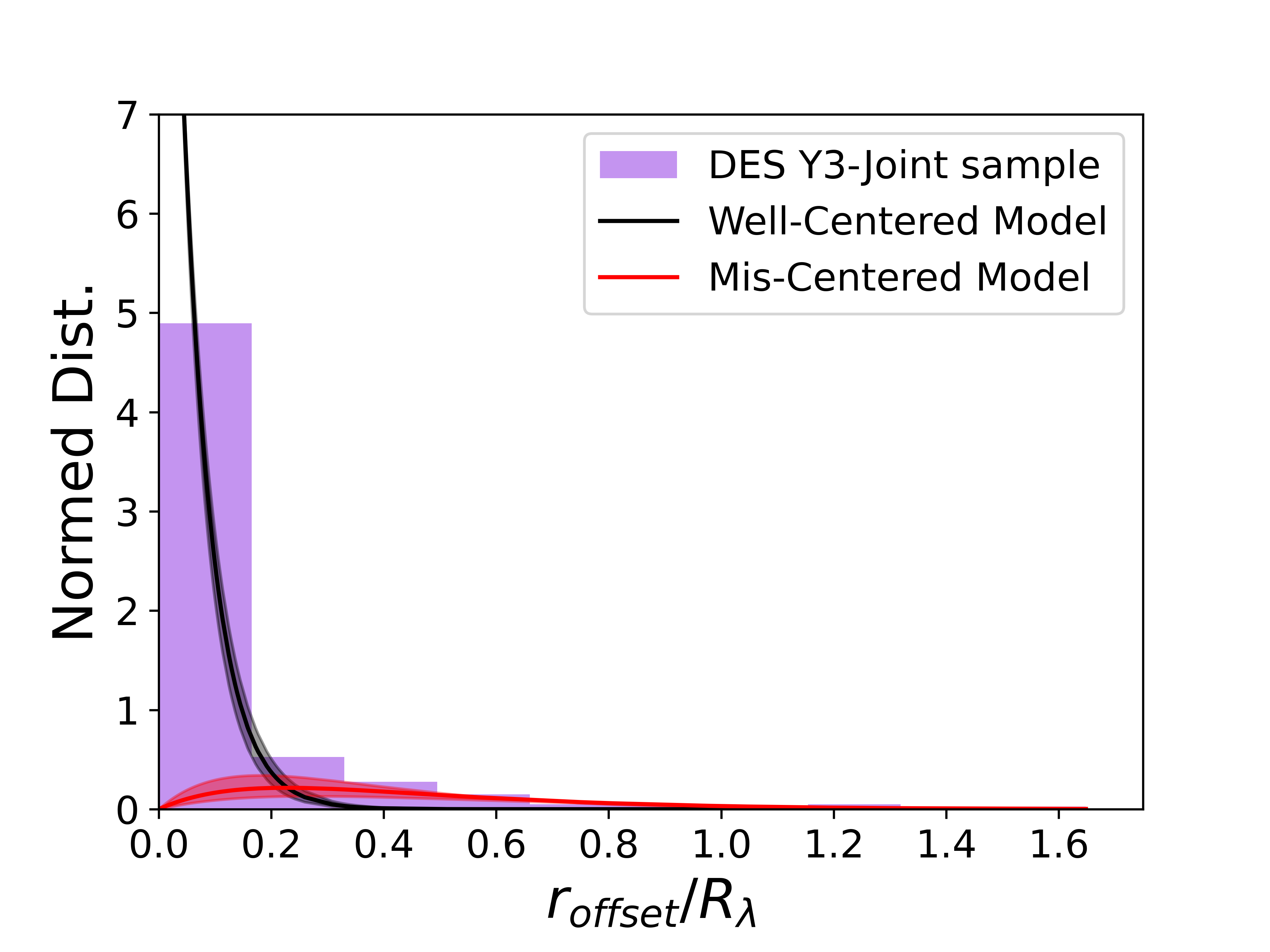}
    \includegraphics[width=8cm]{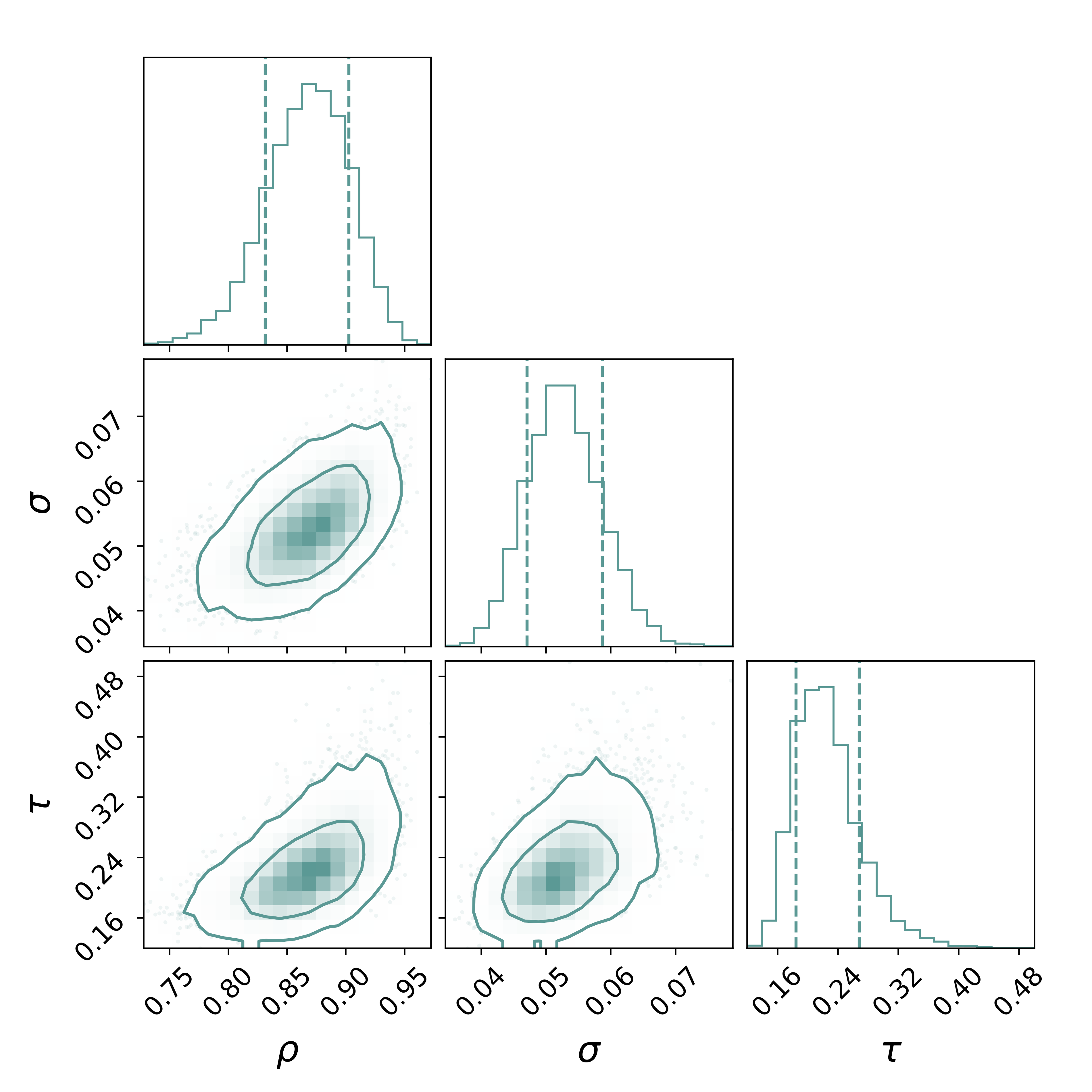}
    \caption{Centering distribution and model fits for the  joint Chandra-XMM sample. \textit{Left:} Histogram of X-ray peak to redMaPPer position offsets with offsets in terms of $R_{\lambda} = (\lambda/100)^{0.2} h^{-1}$ Mpc.  Overlaid are the best fit models and 1$\sigma$ uncertainties for the well centered, $P_\mathrm{cen}$ (black), and miscentered, $P_\mathrm{mis}$ (pink), distributions. \textit{Right:} Parameter constraint distributions for the centering model fit to the joint cluster sample.}
    \label{fig:joint_center}
\end{figure*}

\begin{figure}
    \centering
    \includegraphics[width=8cm]{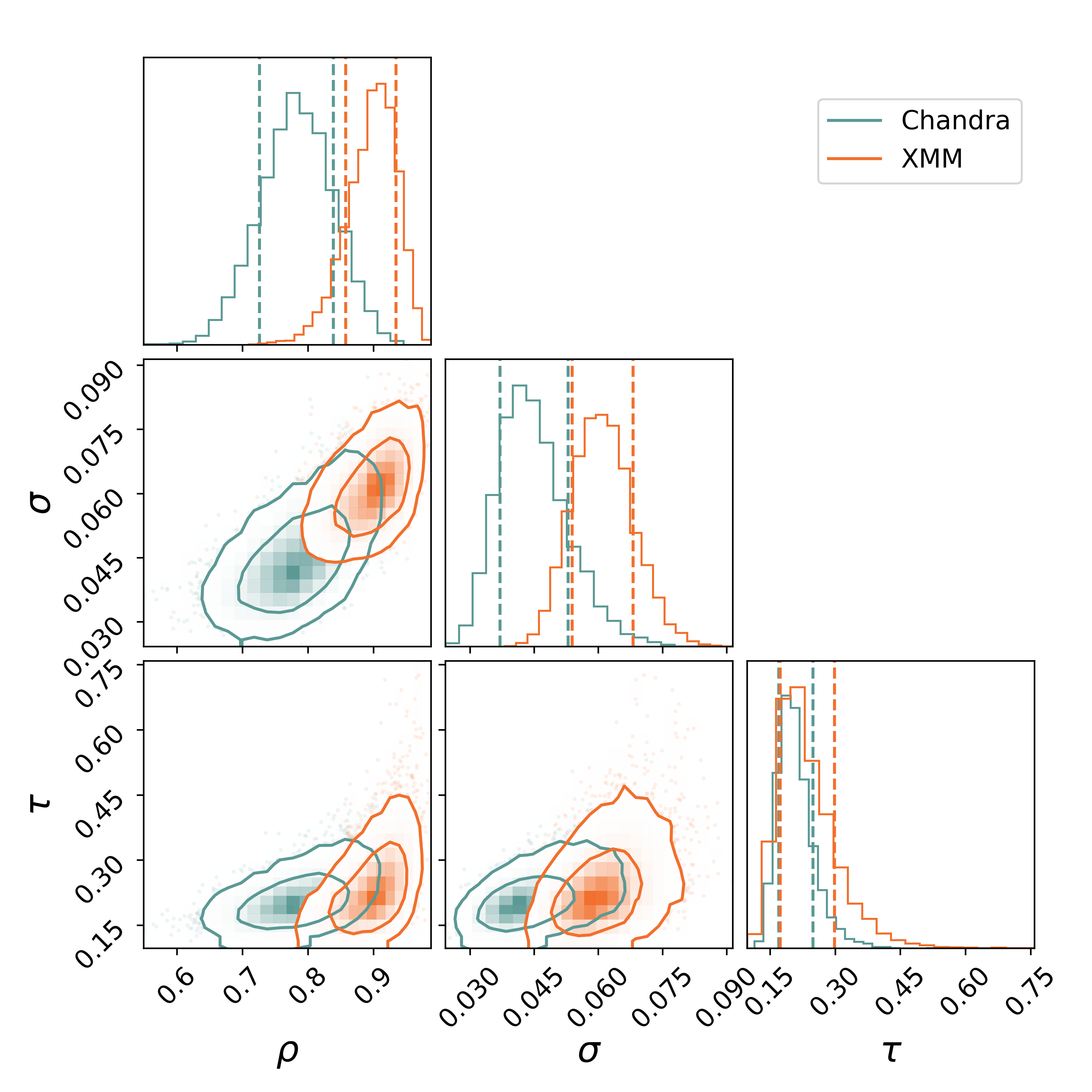}
    \caption{Parameter posterior distributions for the centering model fits to the X-ray peak to redMaPPer position offsets for Chandra (blue) and XMM (orange).}
    \label{fig:chandra_center}
\end{figure}

We first fit the distribution of offsets between the MATCha and XCS determined X-ray peaks and the nominal redMaPPer central galaxy.  Model parameter constraints for the individual Chandra and XMM samples are shown in Table~\ref{tab:cen_summary} and Figure~\ref{fig:chandra_center}.  A histogram of the X-ray peak to redMaPPer offsets is shown in the left panel of Figure~\ref{fig:joint_center}.  While the XMM sample results in a slightly higher well-centered fraction and width of the well-centered distribution, all of the parameters are consistent between the Chandra and XMM fits.  These fits are also consistent with the DES Y1 Chandra and XMM fits in \citet{Zhang19}, but with smaller uncertainties given the larger Y3 sample sizes.

Given the general consistency, we combine the Chandra and XMM samples to give a joint sample of 243 clusters once duplicates are removed.  For duplicate clusters appearing in both samples, we remove the XMM cluster in the joint centering fits given that Chandra has superior spatial resolution.  For the joint fit, we find a well centered fraction $\rho = 0.87 \pm 0.04$ and a width of the miscentered distribution $\tau = 0.23 \pm 0.05$.  The best fit model and distributions of parameter constraints for the joint sample are shown in Figure~\ref{fig:joint_center} and Table~\ref{tab:cen_summary}.

\begin{table}
\begin{center}
\begin{tabular}{|l|c|c|c|c|}
\hline 
Sample & $\rho$ & $\sigma$  & $\tau$ & N\\ \hline
Chandra & 0.78 $\pm$ 0.06  & 0.045 $\pm$ 0.008 & 0.21 $\pm$ 0.04 & 124   \\ 
XMM & 0.90 $\pm$ 0.04 & 0.061 $\pm$ 0.007 & 0.24 $\pm$ 0.07 & 161  \\
Joint   & 0.87 $\pm$ 0.04 & 0.053 $\pm$ 0.006 & 0.23 $\pm$ 0.05 & 243  \\
\hline
\end{tabular}
\caption{Best fit values and 1 $\sigma$ uncertainties on $\rho$, $\sigma$, and $\tau$ for the centering model given by Equation \ref{eq:offset} for the Chandra only, XMM only, and joint cluster samples.}
\label{tab:cen_summary}
\end{center}
\end{table}

\begin{figure}
    \centering
    \includegraphics[width=8cm]{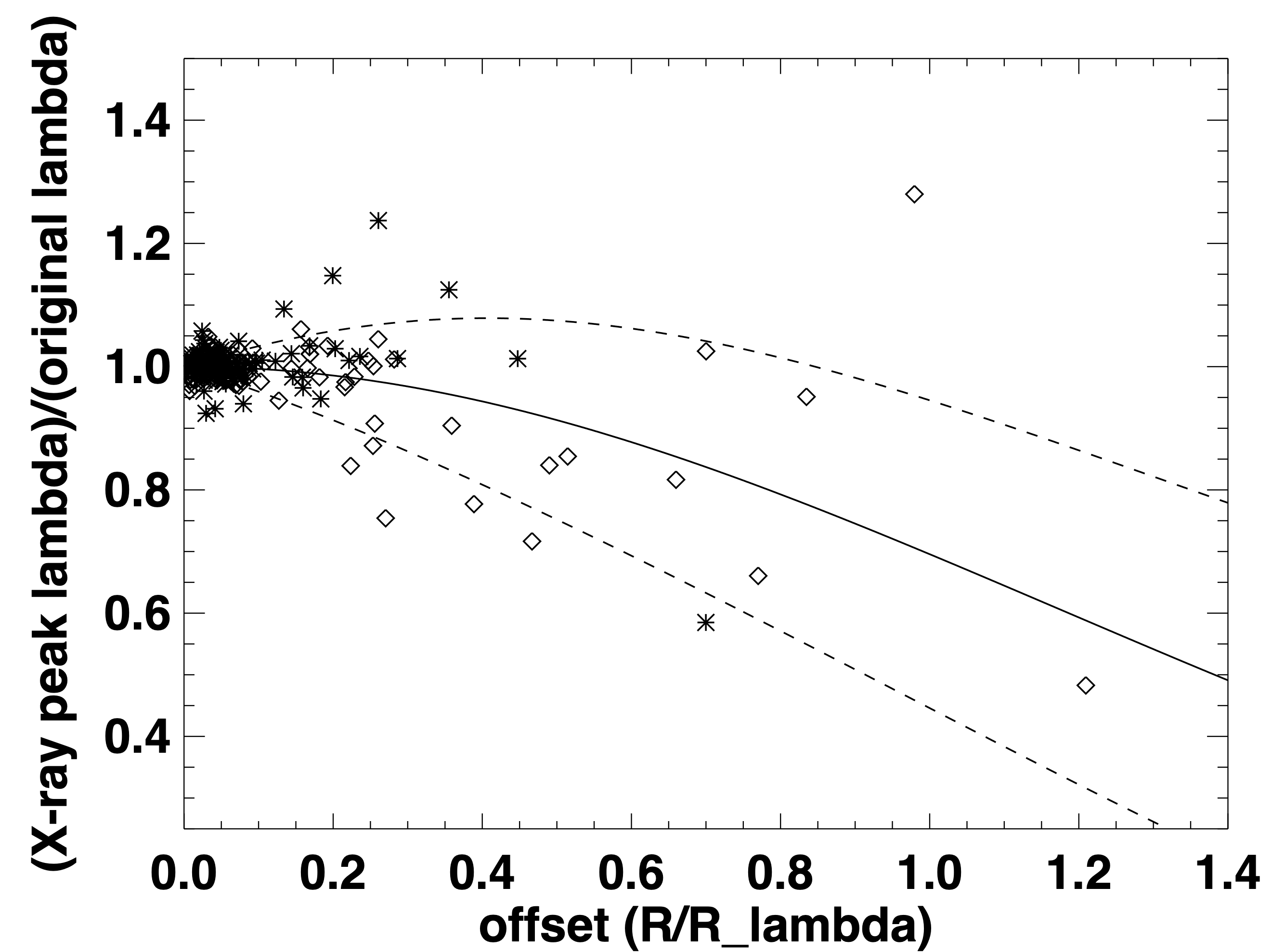}
    \caption{Ratio of the original richness $\lambda$ at the redMaPPer center to the richness calculated at the X-ray peak as a function of X-ray peak to redMaPPer offset distance. Clusters in the Chandra sample are shown with diamonds and in the XMM sample are shown with asterisks; here for clusters common to the two samples the Chandra position is chosen. The solid line shows the model for DES derived in \citet{Zhang19} with dashed lines showing the model 1$\sigma$ dispersion.}
    \label{fig:rich_offset}
\end{figure}

In addition to affecting the stacked lensing signal, miscentering causes a systematic underestimation of cluster richness relative to well-centered clusters which depends on the miscentering offset.  This effect was modeled in \cite{Zhang19} by fitting the ratio of $\lambda$ at the nominal center to $\lambda$ at the position of the second most likely central galaxy as a function of the offset between the two.  We directly probe the richness error by recalculating $\lambda$ at the X-ray peak position for each cluster in our X-ray centering samples. In Figure~\ref{fig:rich_offset}, we show the ratio of $\lambda$ at the X-ray peak to the original redMaPPer $\lambda$ compared to the \cite{Zhang19} model.  The data are broadly consistent with the model, though with somewhat larger scatter.  In general, small offsets of $\sim 20$\% of $R_{\lambda}$ or less lead to little richness bias, while large offsets can lead to significant underestimation of the richness by 40\% or more.

Centering information can also shed light on a particular failure of redMaPPer dubbed \textit{mispercolation} \citep{matcha}. In cases where two spatially close clusters also have similar richnesses, or when redMaPPer has incorrectly split a large system into multiple clusters, redMaPPer can incorrectly assign a smaller richness to a larger system (and vice versa). Visual inspection enables us to identify these errors in the X-ray data. We correct them following the methods outlined in \cite{matcha} and \cite{Zhang19} by manually assigning $L_X$, $T_X$, and the centroids and radii of the brightest X-ray cluster to the richest redMaPPer halo and removing the remaining cluster. In this way, we are able to model the mispercolation as an extreme case of miscentering. In the Chandra sample, we found only one case of mispercolation, and there were none found in the XMM sample.  In general, this failure mode appears to be less frequent for DES than in the SDSS redMaPPer catalog \citep{matcha}.

\subsubsection{Origins of miscentering}

Using the identifications of the correct centrals in miscentered clusters, we explore common reasons for miscentering. The numbers of clusters in the categories identified in Sect. ~\ref{sec:desy3centmethods} for Chandra, XMM, and the joint sample are listed in Table~\ref{tab:centerflags}.  The designation of ``Miscentered" was relatively robust with independently flagged clusters appearing in both the Chandra and XMM samples agreeing in all cases but one, marked miscentered for one telescope and ambiguous for the other; this cluster was subsequently moved to ambiguous. The ``Ambiguous" definition was more subjective with four clusters (two each) being flagged for one telescope and not the other; we have left these flags as is and make no strong conclusions based on the ambiguous clusters.

\begin{table*}
\begin{center}
\begin{tabular}{|l|c|c|c|c|c|}
\hline
Source & Total & Miscentered & Ambiguous & Central & Central not \\
 & Clusters & & & in Top 5 & RM member \\
\hline
Chandra & 124 & 21 & 12 & 7 & 12  \\
XMM   & 161 & 22 & 8 & 11 & 9 \\
Joint & 243 & 34 & 17 & 17 & 14\\
\hline
\end{tabular}
\caption{Visual classification of redMaPPer centering accuracy using the X-ray surface brightness distribution to identify the central galaxy.  Column 2 lists the total number of clusters in each sample; column 3 gives the number of clusters where redMaPPer clearly misidentified the central galaxy, while column 4 lists the number of clusters for which it was ambiguous whether or not the redMaPPer center was correct.  Columns 5 and 6 are subsets of column 3 giving the number of miscentered clusters for which the correct central was one of the possible alternative central galaxies identified by redMaPPer and the number of clusters where the correct central galaxy was not a member of any redMaPPer cluster, respectively.}
\label{tab:centerflags}
\end{center}
\end{table*}

We find that $\sim 14$\% of clusters are miscentered by redMaPPer and an additional $\sim 7$\% are ambiguous.  These fractions are consistent with the miscentered fractions implied by the X-ray peak to redMaPPer offset distributions shown in Table~\ref{tab:cen_summary} and Figure~\ref{fig:joint_center} as are the individual Chandra and XMM fractions. The ambiguous clusters are nearly all merging clusters with multiple substructures each with bright galaxies associated to them.  In these cases the X-ray peak and redMaPPer position sometimes agree while in other cases they pick different substructures, so some fraction of the ambiguous clusters contribute to the X-ray offset distribution.  In terms of cluster cosmology the relevant, but unanswerable, question is which substructure in a merger would the simulation's halo finder choose as the cluster center compared to the redMaPPer choice.  Taking roughly half of the ambiguous fraction is perhaps a good first order estimate of how many of these would be ``miscentered".

For the miscentered clusters, examination of the DES images in many cases reveals the reasons for miscentering.  In a little less than half of the cases, the correct central galaxy is not a member of any redMaPPer cluster, and for two clusters the correct central was designated as a member of a different cluster by redMaPPer.  Of those clusters where the correct central was not a redMaPPer candidate member at all, eight were affected by gaps in the DES data or the presence of a nearby star (in equal proportion) and another one by the presence of a large, very low-redshift galaxy.  In these cases, the central was likely masked out.  An additional two were affected by AGN or star-formation in the central galaxy, and we note that these cases might be over-represented in our archival X-ray samples. The reasons for miscentering in the remaining four clusters in this category are unclear. 

\begin{figure*}
    \centering
    \includegraphics[width=8cm]{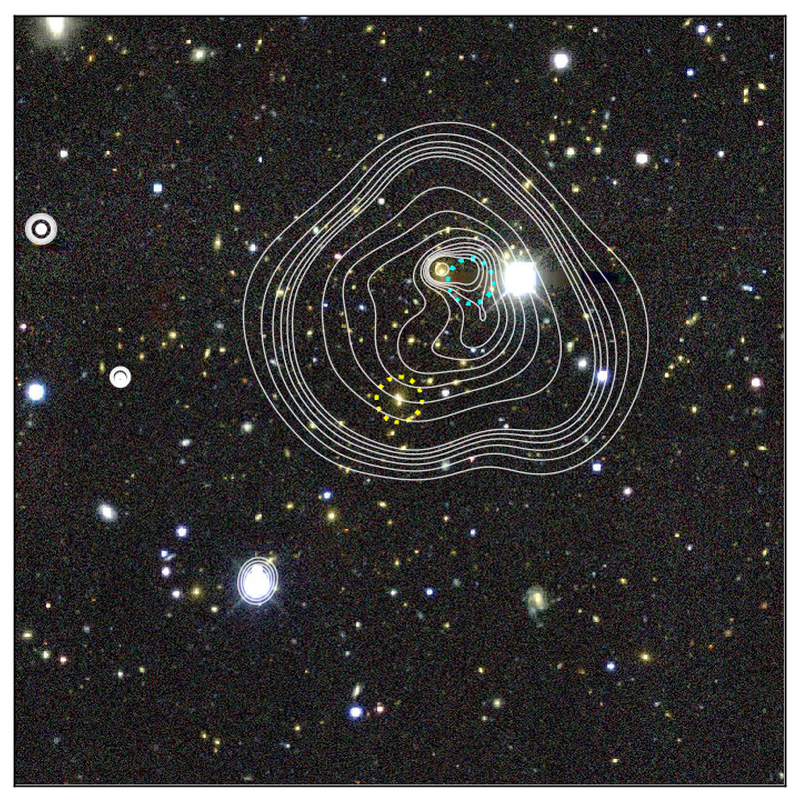}
    \includegraphics[width=8cm]{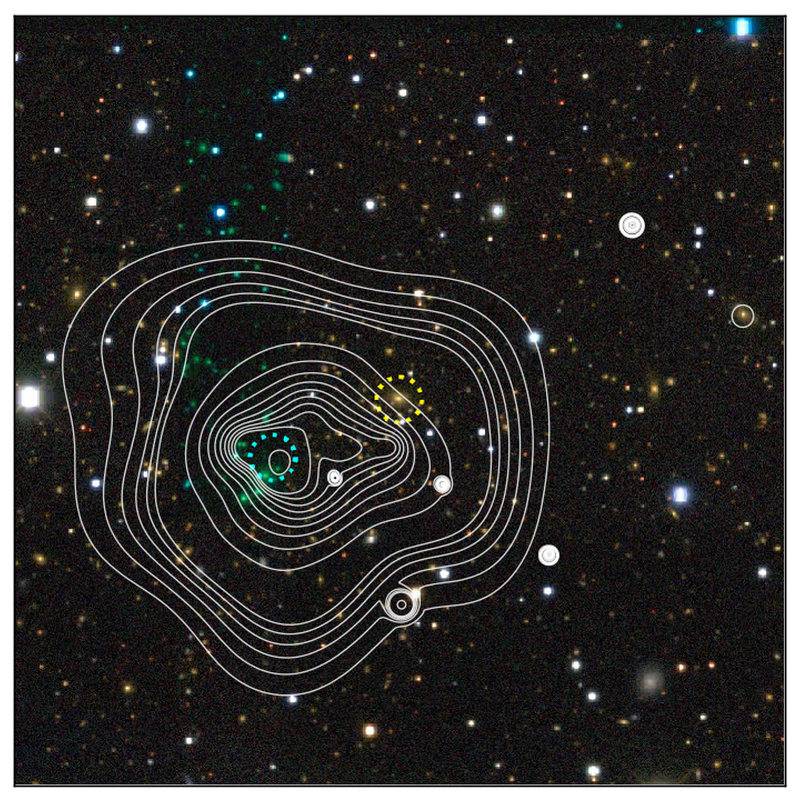}
    \caption{Examples of two clusters where the correct central galaxy was masked in the DES data.  Images are 6x6 arcmin, gri color composites from DES Y3.  Contours show contours of Chandra X-ray brightness.  The yellow dashed circle marks the galaxy chosen by redMaPPer as the central, and the blue dashed circle the X-ray peak location. \textit{Left:} MEM\_MATCH\_ID 403 where the central galaxy is masked by the presence of a bright star. \textit{Right:} MEM\_MATCH\_ID 559 where the central galaxy is masked due to gaps in the data coverage seen as a strip of green/blue galaxies where data is not available for all bands.}
    \label{fig:masked}
\end{figure*}

In half of the miscentered clusters, the correct central galaxy was one of the other possible centrals identified by redMaPPer and predominantly the second most likely. 

\begin{figure}
    \centering
    \includegraphics[width=8cm]{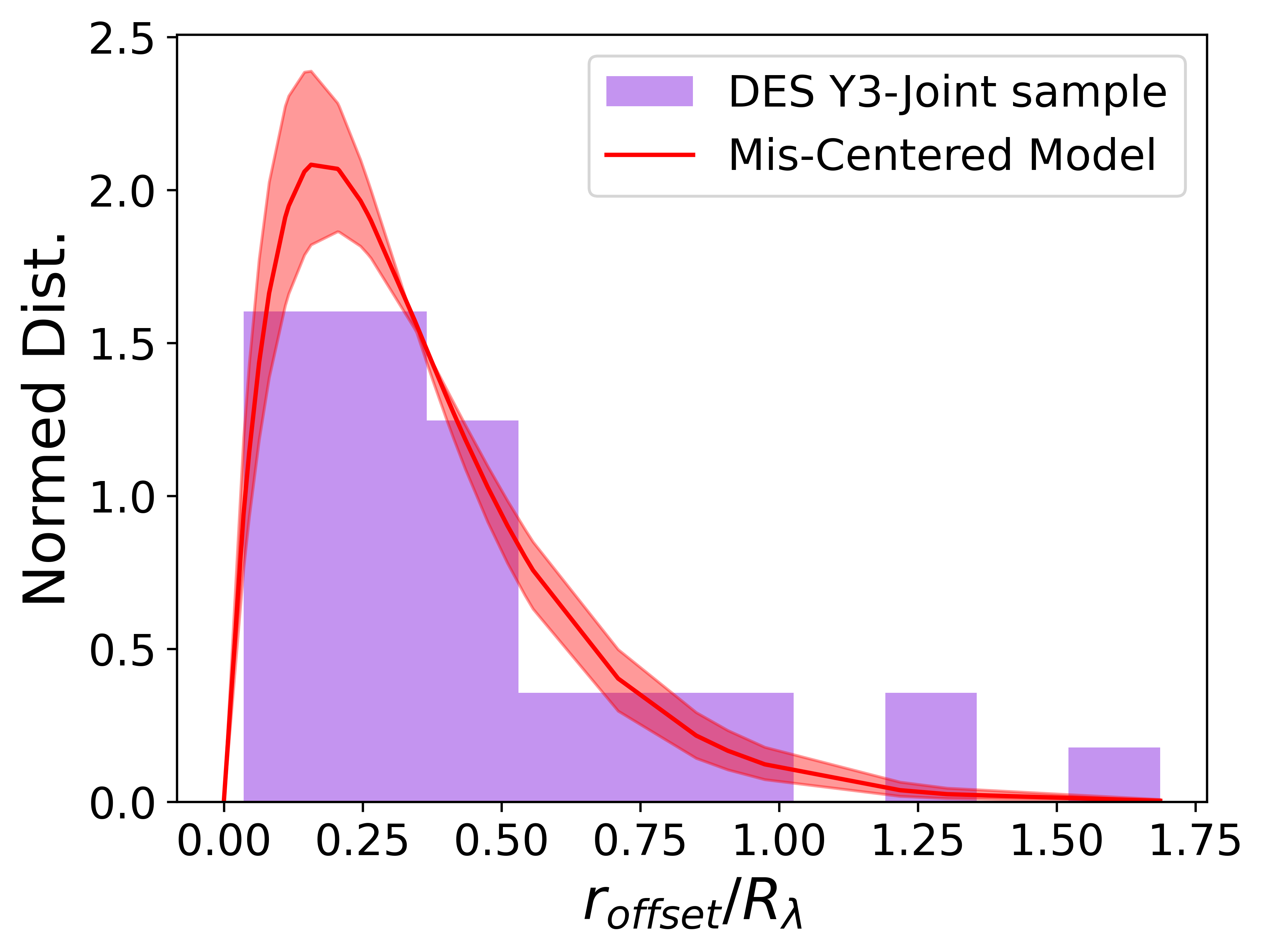}
    \caption{Fit of miscentered clusters only in the joint Chandra-XMM sample. Colors are the same as in the left panel of Figure~\ref{fig:joint_center}.}
    \label{fig:miscenter}
\end{figure}

Figure~\ref{fig:miscenter} shows the distribution of offsets between the redMaPPer chosen central and the true central galaxy for the 34 clusters in the joint sample identified visually as miscenterd.  Fitting the miscentered model $P_\mathrm{mis}$ to this distrubution, we find $\tau = 0.18 \pm 0.02$, slightly smaller but consistent with what we measure for the X-ray to redMaPPer $P_\mathrm{mis}$ model.

\subsubsection{Miscentering Richness and Redshift Trends}

\begin{figure}
    \centering
    \includegraphics[width=8cm]{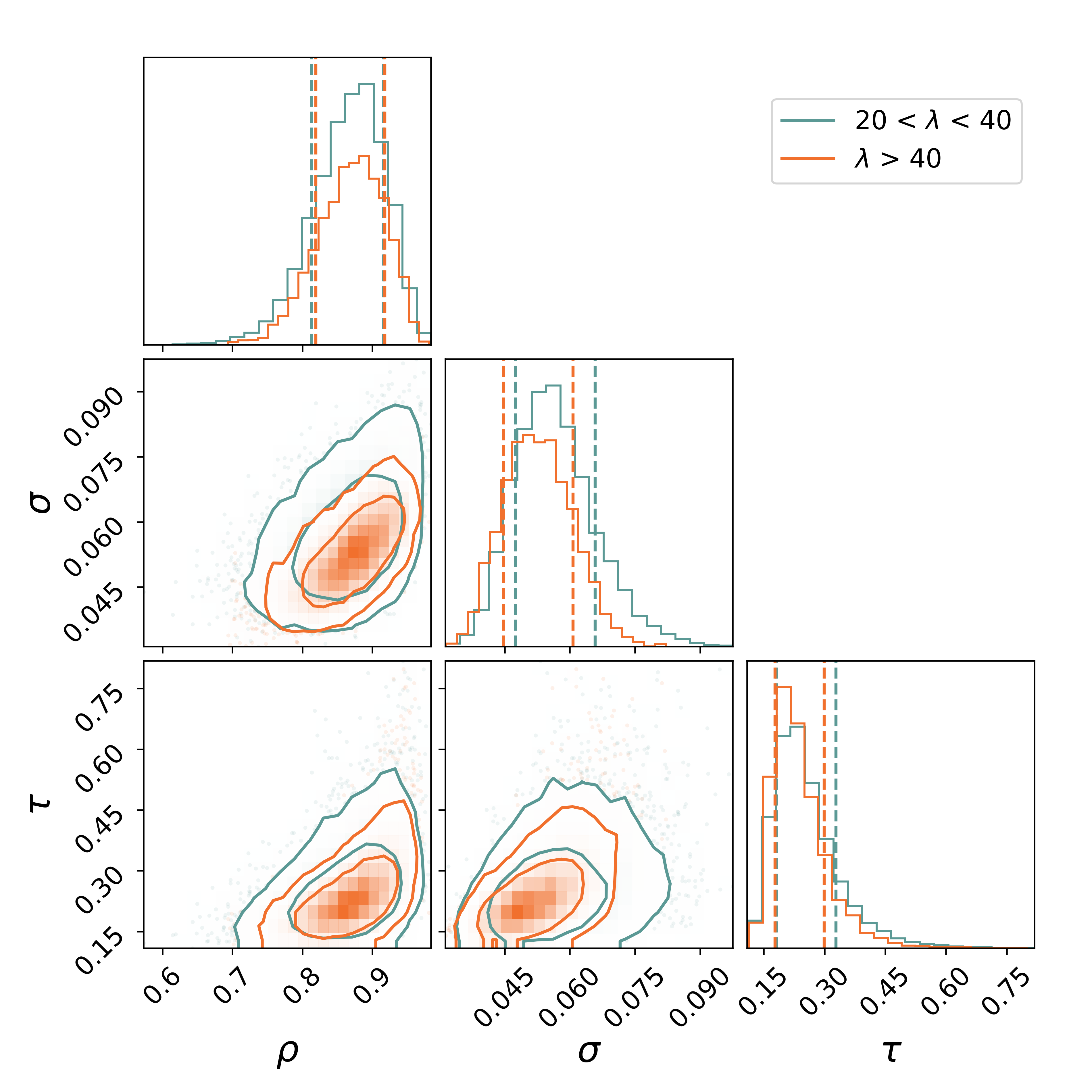}
    \caption{Plots showing the parameter constraint distributions for the joint Chandra  and XMM sample in bins of $\lambda$. The $20 < \lambda < 40$ bin is shown in blue and the $\lambda > 40$ bin is shown in orange.}
    \label{fig:centerlambda40}
\end{figure}

The joint Chandra-XMM centering sample is large enough for us to begin to investigate trends in miscentering across richness and redshift. We first consider bins in richness and cut the joint centering sample on $20<\lambda<40$ (``low-richness'') and $\lambda>40$ (``high-richness''). We do not find significant differences in centering results between our low and high-richness samples (see Table \ref{tab:lambdacenter}, Figure~\ref{fig:centerlambda40}); in fact the two fits are essentially the same and the same as for the full sample. In addition, the range of redshifts and median redshift of the two samples are similar.  A larger miscentering fraction for low-richness clusters might have explained the results of \citet{deskp} which imply the measured lensing signal of low-richness clusters is lower than expected, but this does not seem to be the case in our sample. As our low-richness samples are incomplete, we cannot completely rule out this possibility, but the X-ray undetected, low-richness clusters would need to have significantly worse centering.  We also experimented with a somewhat higher richness cut of $\lambda$ above and below 75 which gives roughly similar numbers of clusters in the two bins, but again found no significant difference (see see Table \ref{tab:lambdacenter}).

\begin{figure}
    \centering
    \includegraphics[width=8cm]{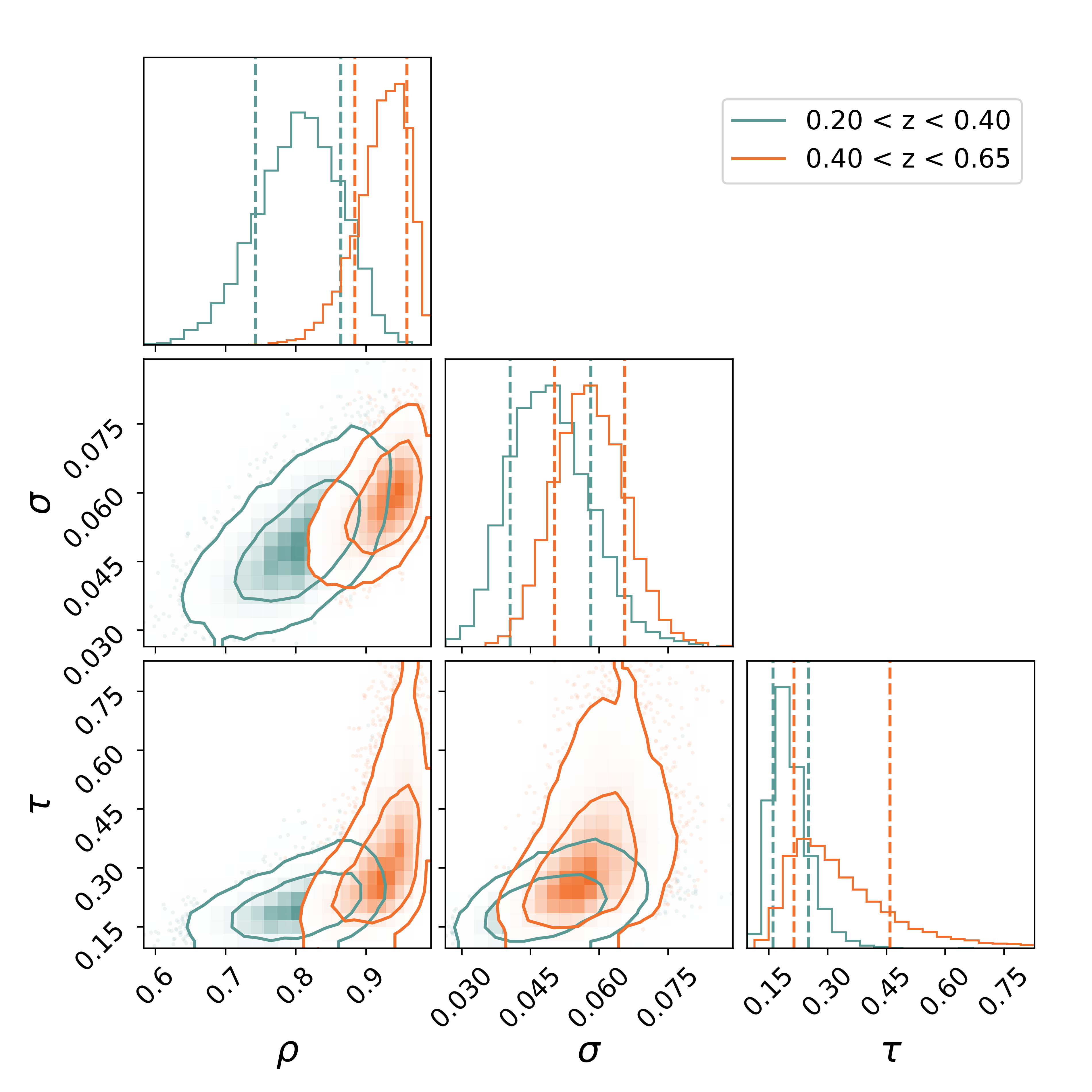}
    \caption{Plots showing the parameter constraint distributions for the joint Chandra  and XMM sample in bins of redshift. The $0.2 < z < 0.4$ bin is shown in blue and the $0.4 < z < 0.65$ bin is shown in orange.}
    \label{fig:centerz}
\end{figure}

To investigate redshift trends, we again use two bins cut on $0.2<z<0.4$ (``low-redshift'') and $0.4<z<0.65$ (``high-redshift''). The results are shown in Figure~\ref{fig:centerz} and Table \ref{tab:zcenter}. We do find a slight increase in the miscentered fraction for the low-redshift sample compared to the high-redshift sample; however, this discrepancy is within $2\sigma$ and coupled with a marginal decrease in the width of the miscentered distribution.  We note that the two samples have very similar median richnesses of 58 and 63, respectively, and similar richness ranges. 

\begin{figure}
    \centering
    \includegraphics[width=8cm]{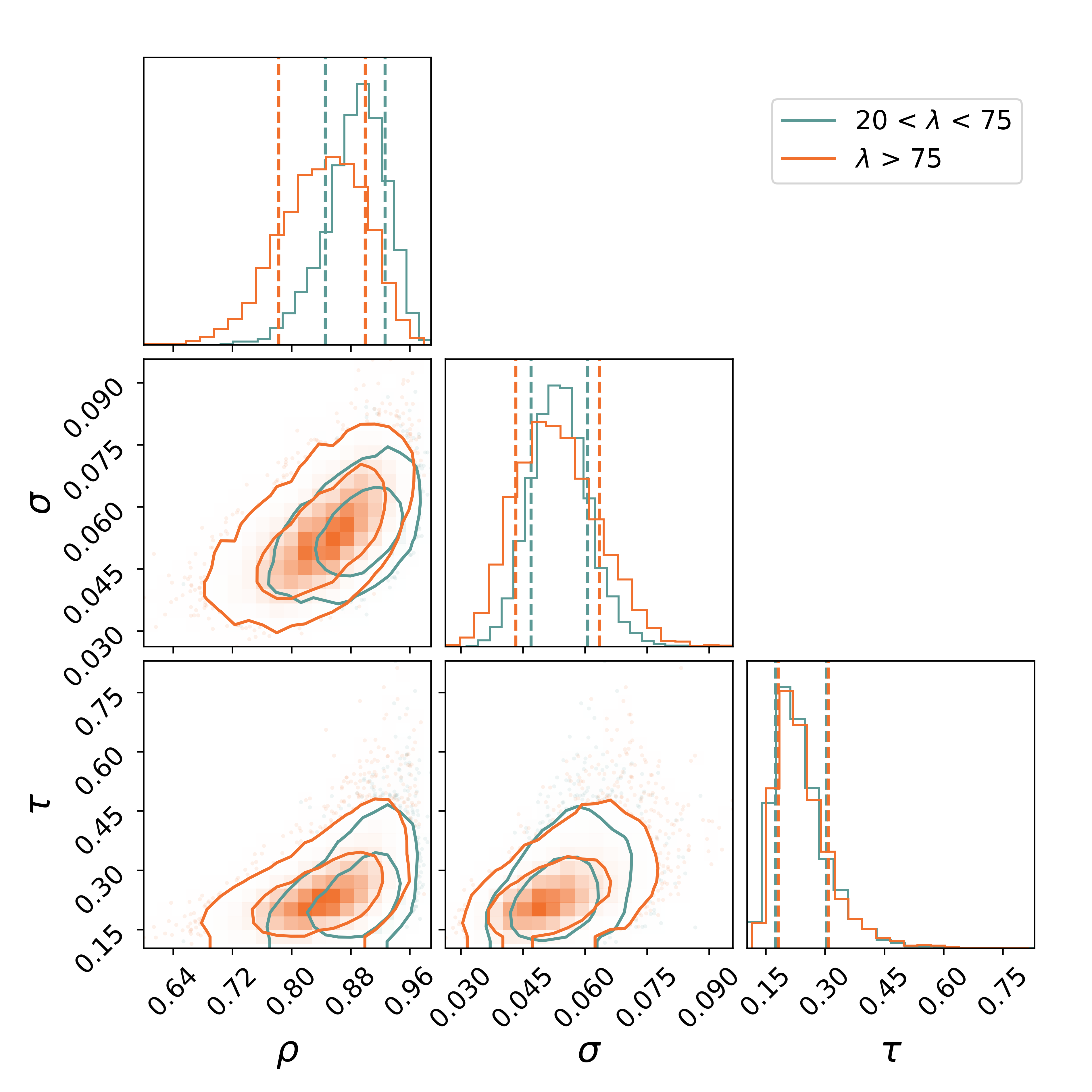}
    \caption{Plots showing the parameter constraint distributions for the joint Chandra  and XMM sample in bins of $\lambda$. The $20 < \lambda < 75$ bin is shown in blue and the $\lambda > 75$ bin is shown in orange.}
    \label{fig:centerlambda75}
\end{figure}

\begin{table}
\begin{center}
\begin{tabular}{|c|c|c|c|c|}
\hline
Sample & $\rho$  & $\sigma$ & $\tau$ & N \\
\hline
$20<\lambda<40$ & $0.86\pm0.05$ & $0.057\pm0.009$ & $0.26\pm0.09$ & 82 \\
$\lambda>40$ & $0.87\pm0.05$ & $0.053\pm0.008$ & $0.24\pm0.07$ & 161\\
$20<\lambda<75$ & $0.89\pm0.04$ & $0.054\pm0.007$ & $0.24\pm0.07$ & 131 \\
$\lambda>75$ & $0.84\pm0.06$ & $0.053\pm0.010$ & $0.25\pm0.08$ & 112 \\
Full joint sample & $0.87\pm0.04$ & $0.053\pm0.006$ & $0.23\pm0.05$ & 243 \\
\hline
\end{tabular}
\caption{Best fit values and $1\sigma$ uncertainties on $\rho$, $\sigma$ and $\tau$ for the centering model given by Equation \eqref{eq:offset} for the low-richness and high-richness samples compared to joint results.}
\label{tab:lambdacenter}
\end{center}
\end{table}

\begin{table}
\begin{center}
\begin{tabular}{|c|c|c|c|c|}
\hline
Sample & $\rho$  & $\sigma$ & $\tau$ & N \\
\hline
$0.2<z<0.4$ & $0.80\pm0.06$ & $0.050\pm0.009$ & $0.21\pm0.05$ & 119  \\
$0.4<z<0.65$ & $0.92\pm0.04$ & $0.058\pm0.008$ & $0.33\pm0.13$ &  124\\
Full joint sample & $0.87\pm0.04$ & $0.053\pm0.006$ & $0.23\pm0.05$ & 243 \\
\hline
\end{tabular}
\caption{Best fit values and $1\sigma$ uncertainties on $\rho$, $\sigma$ and $\tau$ for the centering model given by Equation \eqref{eq:offset} for the low-redshift and high-redshift samples compared to joint results.}
\label{tab:zcenter}
\end{center}
\end{table}

\section{Richness Scaling Relations} \label{scaling}

 In this section, we present results from our regression analysis between X-ray properties such as $T_{\rm X}$, $L_{\rm X}$, and richness. All relations reported in this section are found using an analytical program called CluStR\footnote{https://github.com/sweverett/CluStR} which calls \textit{linmix}, an implementation of the Bayesian regression model introduced in \citep{Kelly}. The model estimates scaling parameters using data augmentation by incorporating heteroscedastic measurement errors, selection effects, and gaussian mixture modeling for the covariates. 

\subsection{Scaling Relation Methods}
\label{sec:desy3scalemethods}

 Our results have the form $\mathrm{ln}(y) = \alpha \mathrm{ln}(\frac{x}{x_{\rm pivot}}) + \beta$, where $\alpha$, $\beta$, and $\sigma$ are the slope, intercept, and lognormal intrinsic scatter, respectively. The temperature and luminosity errors for all fits are represented as symmetric errors, although asymmetric errors are derived from the X-ray fits. Symmetric errors were chosen due to the limitations of $\textit{linmix}$ which implements the regression model \citep{Kelly}. Symmetric errors are calculated, in log space, by averaging the upper and lower errors, and the central value was used as input data point to $\textit{linmix}$. 

In each case, we fit X-ray properties to richness separately for the Chandra and XMM samples and jointly with the functional form given in Equation (\ref{equ:scaling}).

\begin{align}
\label{equ:scaling}
\mathrm{ln}\left(E(z)^{-\frac{2}{3}}kT_{\rm X, r_{2500}}\right) = \alpha  \mathrm{ln}\left(\frac{\lambda_{\rm RM}}{\lambda_{\rm piv}}\right) + \beta
\end{align} 

We set the richness pivot point, $\lambda_{\rm piv}$, to 70 and normalize $T_{\rm X}$ as well as $L_{\rm X}$ by appropriate factors of $E(z)$ 

For the temperature joint fits, we combine the two samples after adjusting to a common temperature scale and removing duplicates.  Here we keep the XMM temperatures for duplicate clusters given their generally smaller uncertainties. From the 98 galaxy clusters used in the Chandra $T_X(r_{2500})$ sample, we removed 34 duplicate clusters that were present in the XMM catalog in the joint XMM and Chandra fit. There were 64 remaining galaxy clusters from the Chandra data set that were added to the 160 clusters in the XMM data set resulting in 224 total clusters used to determine the joint scaling relation. 

To account for the known temperature offset between Chandra and XMM, we use the 34 clusters present in both samples to derive a relation between the temperatures output by MATCha for Chandra observations and those from XCS for XMM observations as run on the DES Y3 samples here. We find, for $r_{2500}$ and $r_{500}$ temperatures respectively, 
 
\begin{align}
\label{equ:Tx-Tx}
\log_{10}(T_{\rm X, r_{2500}}^\mathrm{Chandra}) = 1.01  \log_{10}
(T_{\rm X, r_{2500}}^\mathrm{XMM}) + 0.10
\end{align}

\begin{align}
\label{equ:Tx-Txr500}
\log_{10}(T_{\rm X, r_{500}}^\mathrm{Chandra}) = 1.04  \log_{10}
(T_{\rm X, r_{500}}^\mathrm{XMM}) + 0.09
\end{align}

These results are consistent with the previous relation derived for SDSS redMaPPer clusters in \citet{redmapperSV} despite updates in both algorithms and instrument calibrations and our use of a different fitting method.  Before performing the joint Chandra + XMM $T_{\rm X}$-$\lambda$ fits, the Chandra temperatures are adjusted to the XMM temperature scale. 

 Several $L_{\rm X,r2500}-\lambda_{\rm RM}$ fits were conducted following the form of Equation \eqref{equ:scalingLx}; in all cases we use soft band (0.5-2 keV) luminosity and an $r_{2500}$ aperture. Again, we conducted individual Chandra and XMM fits along with a joint fit. To create the joint sample, duplicate clusters were typically removed from the Chandra sample unless the cluster was detected for Chandra but not XMM. The Chandra and XMM luminosities for detected clusters common to both samples are fairly consistent with the Chandra luminosities being on average 6\% higher; we rescale the Chandra luminosities by this factor in the joint fit.
 
\begin{align}
\label{equ:scalingLx}
{\rm ln}\left(\frac{L_{\rm X, r_{2500}}}{E(z)\cdot10^{44} \rm ergs/s}\right) = \alpha  {\rm ln}\left(\frac{\lambda_{\rm RM}}{\lambda_{\rm piv}}\right) + \beta
\end{align}

\subsection{Scaling Relation Results}
\label{sec:desy3scaleresults}

\begin{figure}
    \centering
    \begin{tabular}{@{}c@{}}
      \includegraphics[width=8cm]{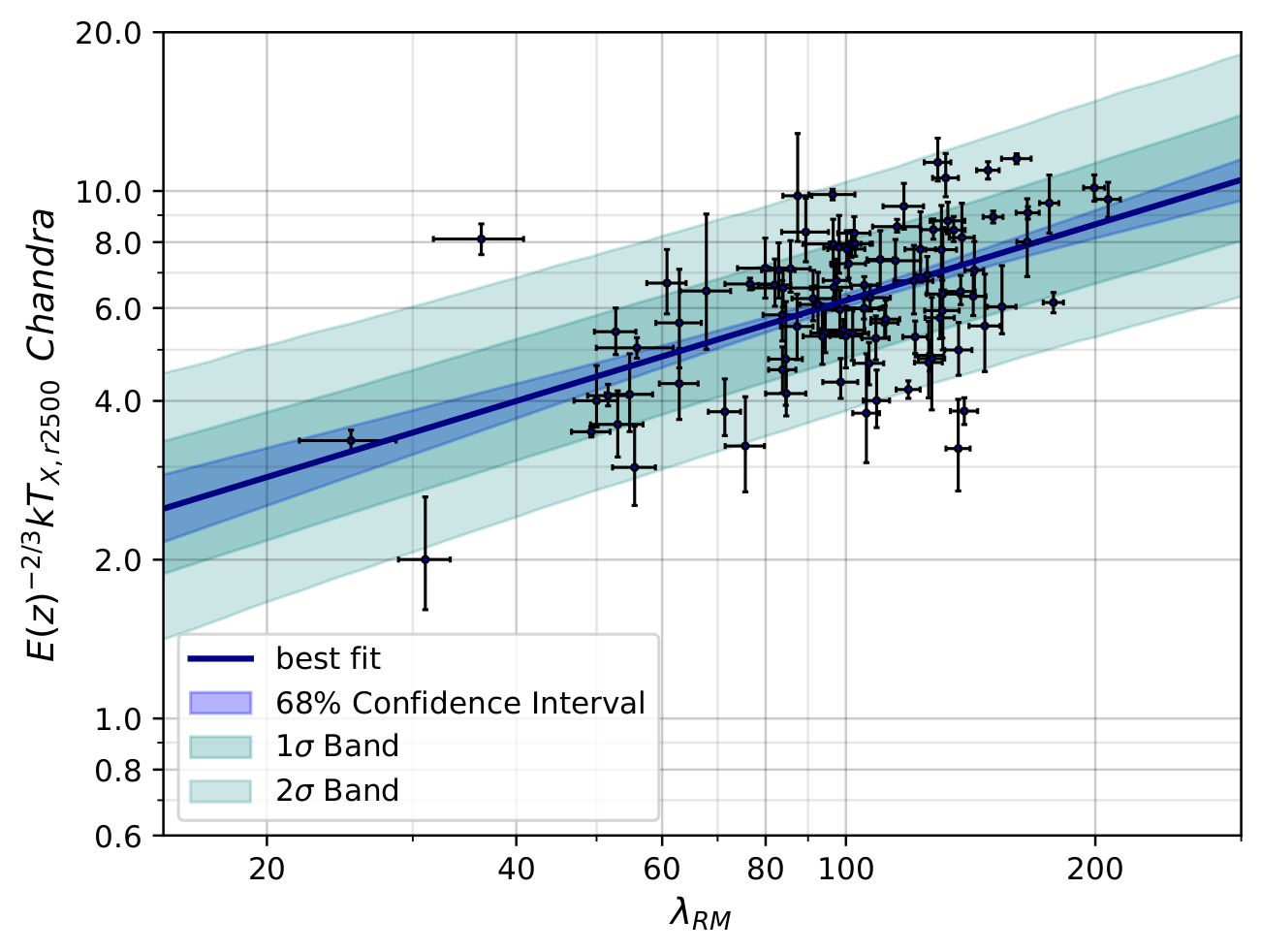}
    \\[\smallskipamount]
    
      \small (a) Chandra

    \end{tabular}
    
    \begin{tabular}{@{}c@{}}
      \includegraphics[width=8cm]{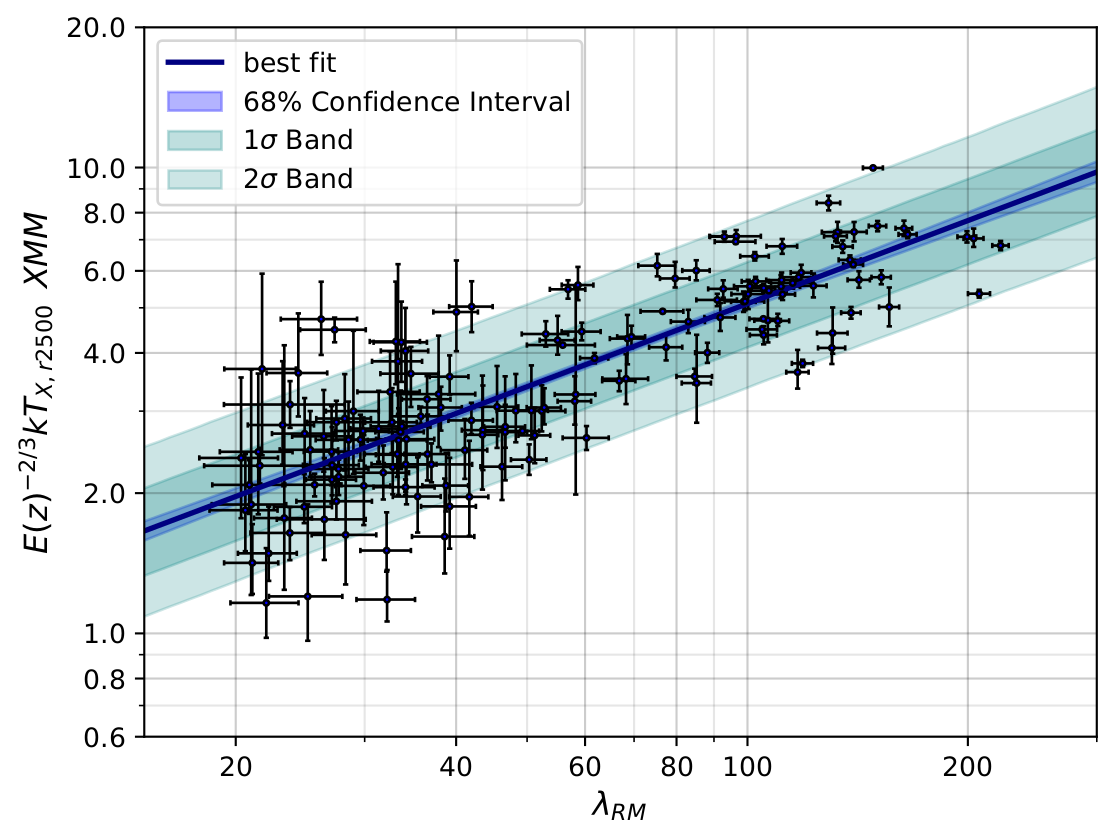} 
    \\[\smallskipamount]
      \small (b) XMM
      
    \end{tabular}
    
    \begin{tabular}{@{}c@{}}
      \includegraphics[width=8cm]{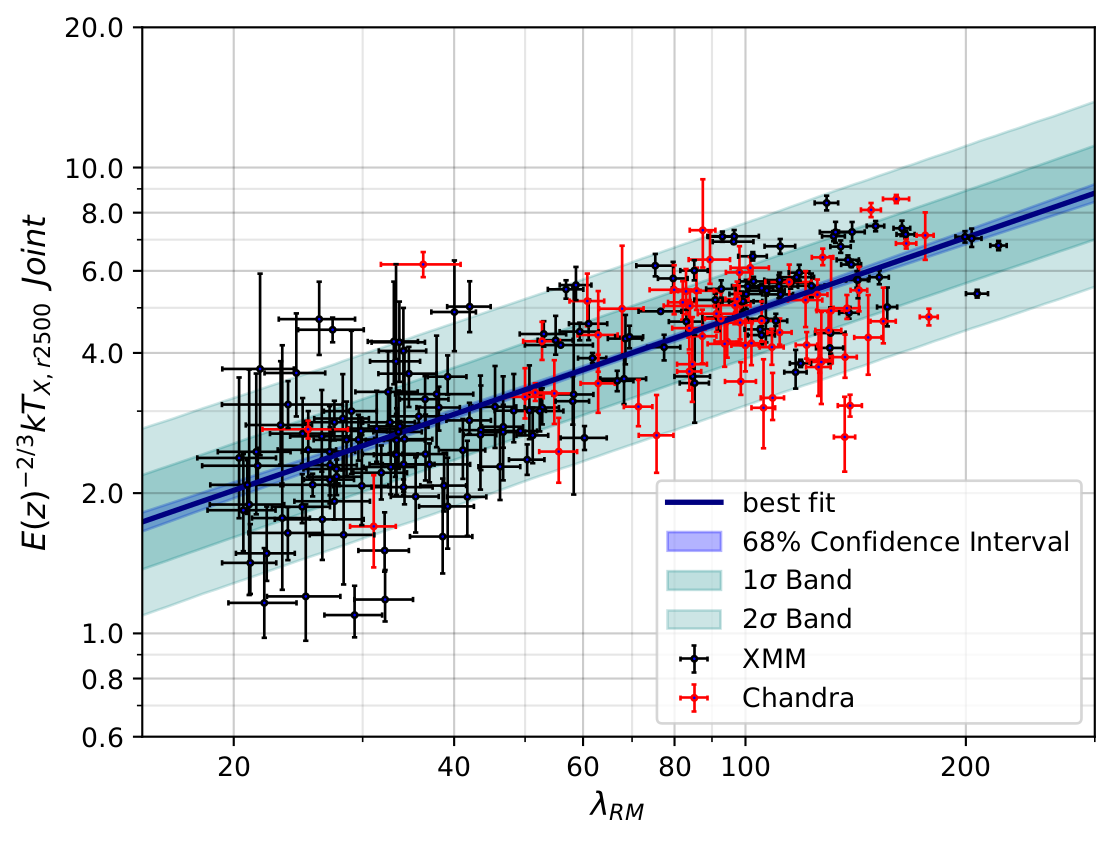} 
      \\[\smallskipamount]
      \small (c) Joint
    \end{tabular}
    
    \caption{$T_{\rm X}(r_{\rm 2500})-\lambda$ plots for a) Chandra,  b) XMM, and c) joint Chandra and XMM samples. For the individual fits, the black points along with their associated errors represent the galaxy clusters from the corresponding data set. For the joint fit, the red points represent the clusters from Chandra that were not already cataloged by the XMM sample. The dark blue line denotes the best-fit, with the richness pivot point set to 70 with the lighter shade of blue representing the 68$\%$ confidence intervals on the fit. The 1$\sigma$ and 2$\sigma$ scatter are shown in dark green and light green respectively. In the joint fit, Chandra temperatures have been adjusted to the XMM temperature scale using Equation~\eqref{equ:Tx-Tx}.} \label{fig:scalingrelations}
\end{figure}

Results for r$_{2500}$ $T_{\rm X}$-$\lambda_{\rm RM}$ relation with the cuts mentioned in Section \ref{Chandra} are reported in Table~\ref{tab:fitr2500} and are shown in Figure~\ref{fig:scalingrelations}. The slope and scatter of the $T_X-\lambda$ relations for the individual Chandra and XMM samples are consistent, with the Chandra relation having a somewhat shallower slope likely due to the lack of many low-richness clusters in this sample.  The normalization of the Chandra relations is higher, showing the known offset between Chandra and XMM temperature estimates \citep{Schell15}.
 
Given the general consistency, we combined the two samples to perform a joint Chandra + XMM $T_{\rm X}$-$\lambda$ fit in an $r_{2500}$ aperture. We found the slope to be $0.54 \pm 0.03$ which is consistent with the slope of $0.62 \pm 0.04$ found in \citet{farahi19} for DES Y1. We find the scatter to be $0.22 \pm 0.01$ while \citet{farahi19} found a somewhat larger scatter of $0.275 \pm 0.019$. 
  
As discussed in Section~\ref{sec:desy3cent}, the redMaPPer estimated richnesses can be biased low for miscentered clusters. Therefore, we refit the $T_{X}-\lambda$ relation using the richnesses calculated centered on the X-ray peak after removing clusters with significant percolation\footnote{redMaPPer performs a percolation step on clusters close in proximity to avoid double counting galaxies. This step is not implemented in X-ray centered runs where the presence of other clusters is effectively not known.}.  Table~\ref{tab:fitr2500} gives the results for the joint Chandra and XMM fit for an $r_{2500}$ aperture for which the total sample was 215 unique clusters.  This fit is very similar to the previous one using the nominal redMaPPer $\lambda$.

The scaling relations for the temperature within an $r_{500}$ aperture are reported in Table \ref{tab:fitr500}; these results are generally in agreement with those found by \citet{Upsdell23} for clusters in the XXL and other XMM survey regions despite the use of a different fitting algorithm.  However, since many of the clusters studied in \cite{Upsdell23} are in common with the clusters contained in this work (since they are also derived from the DES Y3 sample), the agreement with \cite{Upsdell23} is expected.  We therefore compare to the $T_{\rm X}$-$\lambda$ relation presented in \cite{Giles}, constructed from SDSS redMaPPer clusters with available XMM data.  The slope of the relation in \cite{Giles} is in very good agreement with the results reported in Table \ref{tab:fitr500}.  Finally, we compare to results independent of redMaPPer.  \citet{Oguri18} constructed a sample of clusters using the CAMIRA algorithm run on Hyper-Suprime Cam (HSC) observations. \citet{Oguri18} investigated the form of the $T_{\rm X}$-richness relation using richnesses estimated from CAMIRA and XMM temperatures.  They found a slope of 0.50$\pm$0.12, again consistent with the results given in Table \ref{tab:fitr500}.

The $L_{\rm X,r2500}-\lambda_{\rm RM}$ fits were found using soft band (0.5-2 keV) luminosity and an $r_{2500}$ aperture. The results for these fits are reported in Table~\ref{tab:fitlr2500}. We observe that slope of the Chandra sample of 1.36 $\pm$ 0.16 is significantly shallower than the slope of the XMM sample of 1.95 $\pm$ 0.10. The shallower Chandra slope is similar to that seen for the temperature fits and again could be because there are relatively few low-richness clusters in the Chandra sample.  The Chandra $L_{\rm X} - \lambda$ relation for detected clusters is very similar the one found by \cite{matcha} for SDSS redMaPPer clusters.
 
 For the joint Chandra + XMM $L_{\rm X}$-$\lambda$ fit, we find a slope of $1.86 \pm 0.09$, consistent with the XMM result, and a scatter of $0.82 \pm 0.04$.  The $L_{\rm X}-\lambda_{\rm RM}$ relation considering only detected clusters is prone to selection bias.  As a test of the sensitivity of our results, we also fit the $L_{\rm X}-\lambda_{\rm RM}$ relation including luminosity upper limits on undetected clusters as censored data \citep{Kelly}; specifically, we utilize the 3$\sigma$ upper limits on the luminosity for undetected clusters, and for detected clusters with SNR$<9$ we take the upper limit on the measured luminosity as the censored data point.
 Including upper limits, we get a joint Chandra and XMM sample of 676 clusters for which we find a somewhat shallower $L_{\rm X}-\lambda_{\rm RM}$ slope of $1.57 \pm 0.06$ and a marginally larger scatter of $0.88 \pm 0.02$.


\begin{table}
\begin{center}
\begin{tabular}{|l|c|c|c|c|c|}
\hline
Sample & $\alpha$ & $\beta$ & $\sigma$ & N  \\ \hline
Chandra & $0.48 \pm 0.08$ & $1.66 \pm 0.04$ &  $0.25 \pm 0.02$ & $98$\\
XMM & $0.59 \pm 0.03$ & $1.42 \pm 0.02$ & $0.21^{+0.02}_{-0.01}$ & $160$ \\
Joint & $0.54 \pm 0.03$ & $1.39 \pm 0.02$ & $0.22 \pm 0.01$ & $224$ \\ 
Joint ($\lambda_{\mathrm{X-ray}}$) & $0.56 \pm 0.03$ & $1.38 \pm 0.02$ & $0.22 \pm 0.01$ & $215$\\ 
\hline
\end{tabular}
\caption{Parameters for $T_{\rm X}(r_{2500}) - \lambda$ scaling relations. The richness pivot point, $\lambda_{\rm piv}$, was set to 70.}
\label{tab:fitr2500}
\end{center}
\end{table}

\begin{table}
\begin{center}
\begin{tabular}{|l|c|c|c|c|}
\hline
Sample & $\alpha$ & $\beta$ & $\sigma$ & N  \\ \hline
Chandra & $0.43 \pm 0.09$ & $1.72\pm 0.05$ &  $0.27 \pm 0.03$ & $96$\\
XMM  & $0.61 \pm 0.04$ & $1.39 \pm 0.02$ &  $0.23 \pm 0.02$ & $148$\\ 
Joint  & $0.56 \pm 0.03$ & $1.38 \pm 0.02$ &  $0.25 \pm 0.02$  & $210$ \\ \hline
\end{tabular}
\caption{Parameters for $T_{\rm X}(r_{500}) - \lambda$ scaling relations. The richness pivot point, $\lambda_{\rm piv}$, was set to 70. }
\label{tab:fitr500}
\end{center}
\end{table}

\begin{table}
\begin{center}
\begin{tabular}{|l|c|c|c|c|}
\hline
Sample & $\alpha$ & $\beta$ & $\sigma$ & N  \\ \hline
Chandra & $1.36 \pm 0.16$ & $0.04 \pm 0.08$ & $0.76 \pm 0.06$ & 113\\
XMM  & $1.95 \pm 0.10$ & $-0.38 \pm 0.07$ &  $0.83 \pm 0.05$ & 165\\ 
Joint  & $1.86 \pm 0.09$ & $-0.35 \pm 0.05$ &  $0.82 \pm 0.04$  & 239\\ 
+ upper limits  & $1.57 \pm 0.06$ & $-0.38 \pm 0.05$ &  $0.88 \pm 0.02$  & 676\\ 
\hline
\end{tabular}
\caption{Parameters for $L_{\rm X}(r_{2500}) - \lambda$ scaling relations. The richness pivot point, $\lambda_{\rm piv}$, was set to 70.} 
\label{tab:fitlr2500}
\end{center}
\end{table}

\begin{figure}
    \centering
    \begin{tabular}{@{}c@{}}
      \includegraphics[width=8cm]{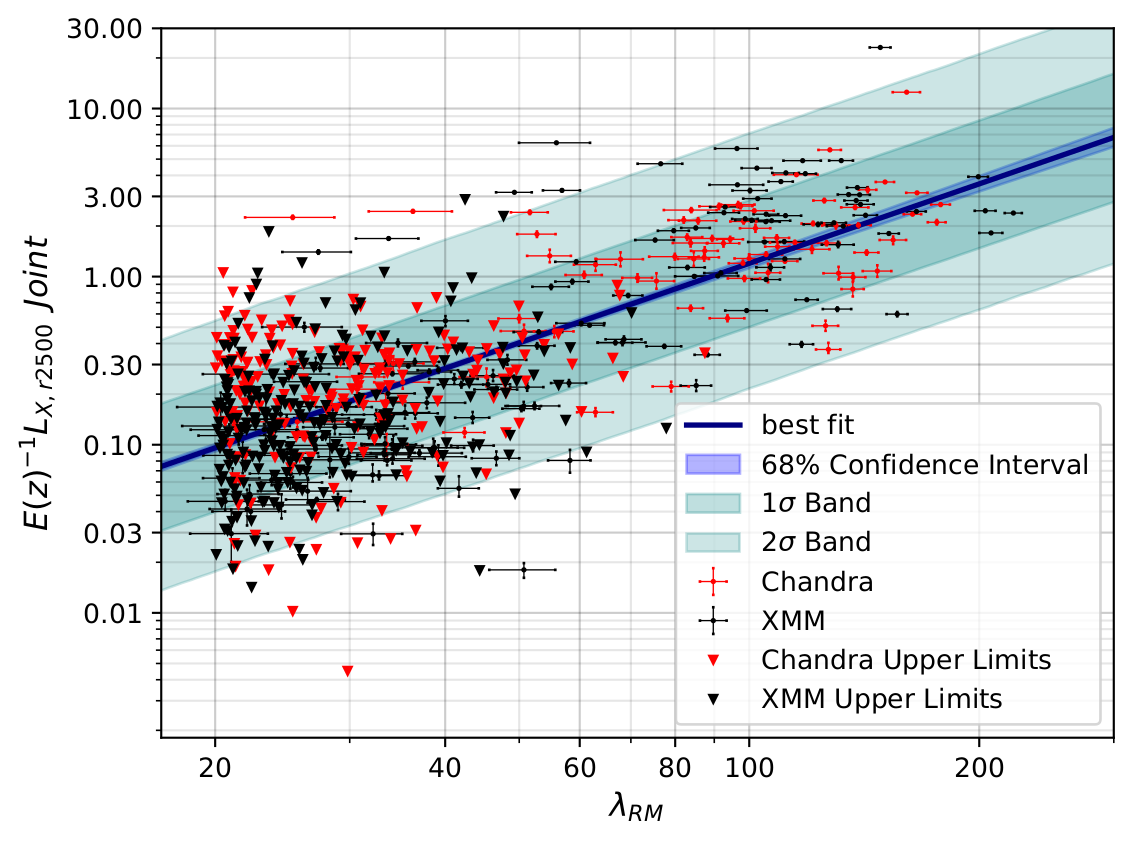}
    \end{tabular}
    \caption{$L_{\rm X}(r_{\rm 2500})-\lambda$ plot for the joint Chandra and XMM data displayed in red and black, respectively. The repeat observations were examined individually, but generally we chose to keep points from the XMM data. For undetected clusters, the 3 $\sigma$ upper limits are displayed with carets. The best-fit line is shown in dark blue; the lighter shade of blue represents the 68$\%$ confidence interval on the fit. The 1$\sigma$ and 2$\sigma$ scatter regions are shown in dark green and light green, respectively. Chandra observations were scaled by 0.94 to match XMM based on the percent difference between the 39 repeat observations. }\label{fig:luminosityupper}
\end{figure}

\subsubsection{Redshift Trends}

\begin{figure}
    \centering
    \begin{tabular}{@{}c@{}}
    
      \includegraphics[width=8cm]{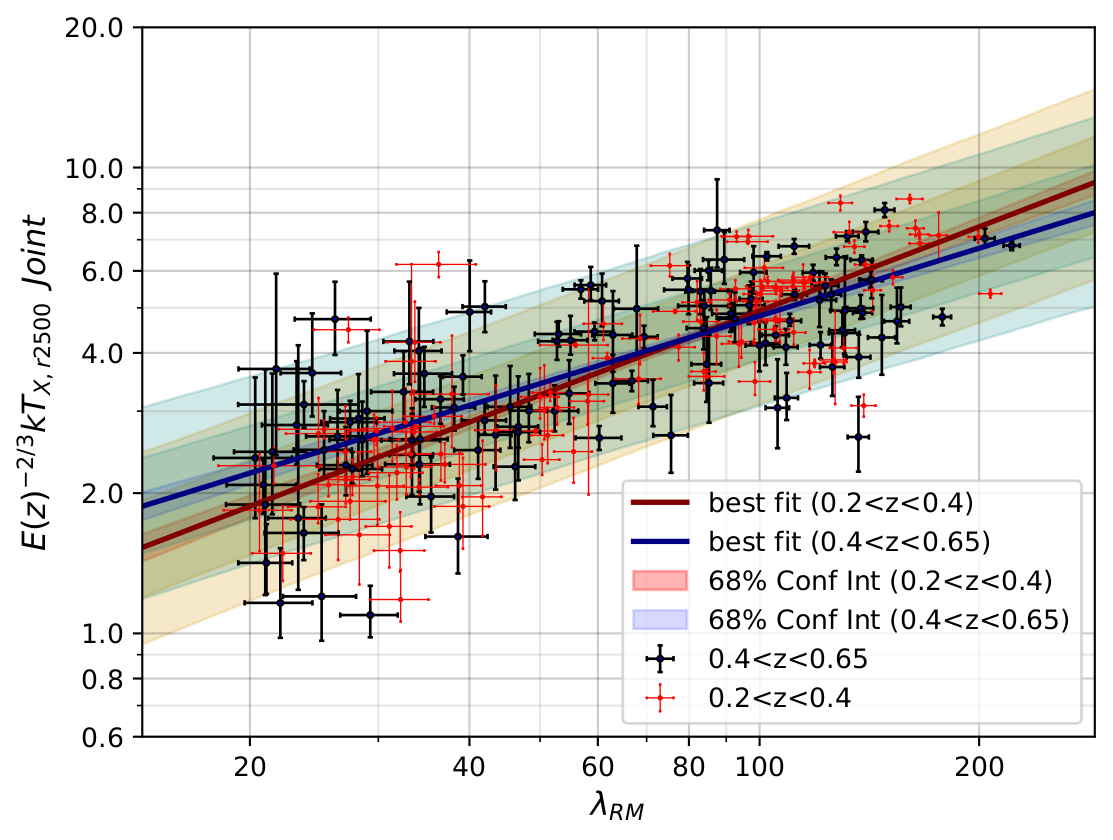}
    \end{tabular}
    
    \begin{tabular}{@{}c@{}}
      \includegraphics[width=8cm]{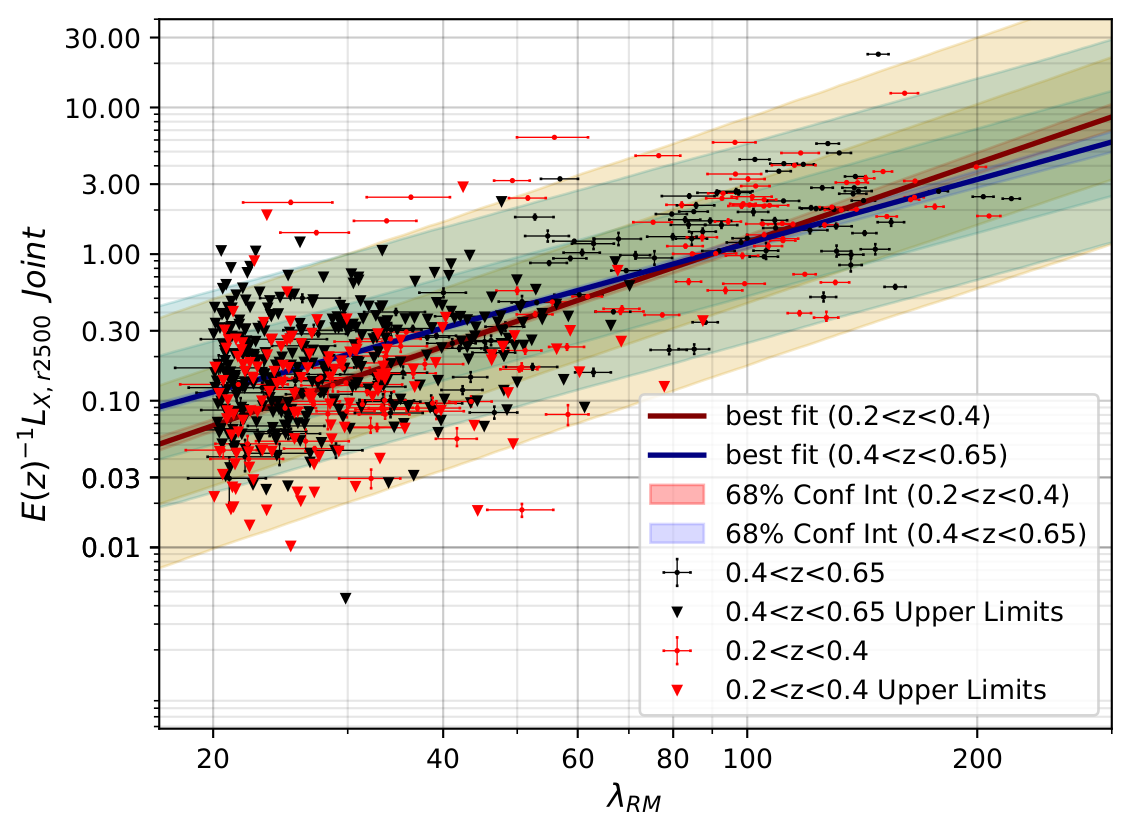}
    \end{tabular}
    
    \caption{X-ray - richness scaling relations in bins of redshift for $T_{\rm X}(r_{\rm 2500})-\lambda$ (top) and $L_{\rm X}(r_{\rm 2500})-\lambda$ (bottom). Red data points and line show clusters in the $0.2 < z < 0.4$ bin and the corresponding best-fit model. The yellow shaded regions show the 1 and 2$\sigma$ scatter for the lower-redshift bin. Black data points and the blue line show clusters in the $0.4 < z < 0.65$ bin and the best-fit model with green shaded regions showing the 1 and 2$\sigma$ scatter.}  \label{fig:redshiftranges}
\end{figure}

We next explore whether there is evidence of a redshift dependence in the scaling relations within the redshift range adopted for DES cluster cosmology, $0.2 < z < 0.65$.  We divide clusters into two, non-overlapping redshift bins of 0.2 $<$ z $<$ 0.4 and 0.4 $\leq$ z $<$ 0.65 and refit the X-ray scaling relations. These redshift cuts were chosen to ensure that we had a similar and substantial sample size on both sides. The results can be found in Table~\ref{table:redshift} and are shown in Figure~\ref{fig:redshiftranges}.

For the $T_{\rm X}-\lambda$ relation, we found the scatters were consistent across all redshift bins, but not all of the slopes were. 
We find that our slope value of $0.60\pm 0.04$ from the redshift range $0.2 < z < 0.4$ is 
larger than our slope value of 0.48 $\pm$ 0.04 from the redshift range $0.4 \leq z < 0.65$. The two samples have roughly similar median richnesses of 75 and 65, respectively, and similar richness ranges. Comparing our results to \cite{farahi19}, we find a similar slope in the low-redshift bin compared to their reported slope of $0.59 \pm 0.05$, but a shallower slope in the high-redshift bin compared to the $0.65 \pm 0.06$ reported in \cite{farahi19}.


We observe the same trend of decreasing slope with increasing redshift in the $L_{\rm X}$-$\lambda$ relation including upper limits. This can be seen in our slope value of 1.79 $\pm$ 0.10 for our 0.2 $<$ z $<$ 0.4 bin compared to 
our slope value of 1.45 $\pm$ 0.07 for the 0.4 $<$ z $<$ 0.65 bin.  We also find a marginally significant decrease in the scatter for the high-redshift bin compared to the low-redshift bin. 
Comparing our results to \citet{matcha}, we see that our low-redshift slope is consistent with their reported slope of 1.78 $\pm$ 0.12 for their redshift range of 0.1 $<$ z $<$ 0.35.

\begin{table*}
\begin{center}
\begin{tabular}{l c c c c }
\hline
Relation & $\alpha$ & $\beta$ & $\sigma$  & $N$ \\ \hline
$T_{\rm X}$-$\lambda$ (0.2 $<$ z $<$ 0.4) & $0.60 \pm 0.04$ & $1.38 \pm 0.02$ & $0.23 \pm 0.02$ & $109$\\
$T_{\rm X}$-$\lambda$ (0.4 $\leq$ z $<$ 0.65) & $0.48 \pm 0.04$ & $1.40 \pm 0.02$ & $0.23 \pm 0.02$ & $116$\\ \hline
$L_{\rm X}$-$\lambda$ (0.2 $<$ z $<$ 0.4)  & $1.79 \pm 0.10$ & $-0.45 \pm 0.08$ & $0.95 \pm 0.05$ & $248$\\
$L_{\rm X}$-$\lambda$ (0.4 $\leq$ z $<$ 0.65)  & $1.45 \pm 0.07$ & $-0.35 \pm 0.07$ & $0.81 \pm 0.03$ & $428$\\ \hline
\end{tabular}
\caption{$T_{\rm X} (r_{2500})-\lambda$ and $L_{\rm X} (r_{2500})-\lambda$ relations in different redshift bins; all fits use the joint Chandra and XMM sample, and the luminosity fit includes upper limits for undetected clusters.}
\label{table:redshift}
\end{center}
\end{table*}

\subsubsection{Richness Trends}
\begin{figure}
    \centering
    \begin{tabular}{@{}c@{}}
      \includegraphics[width=8cm]{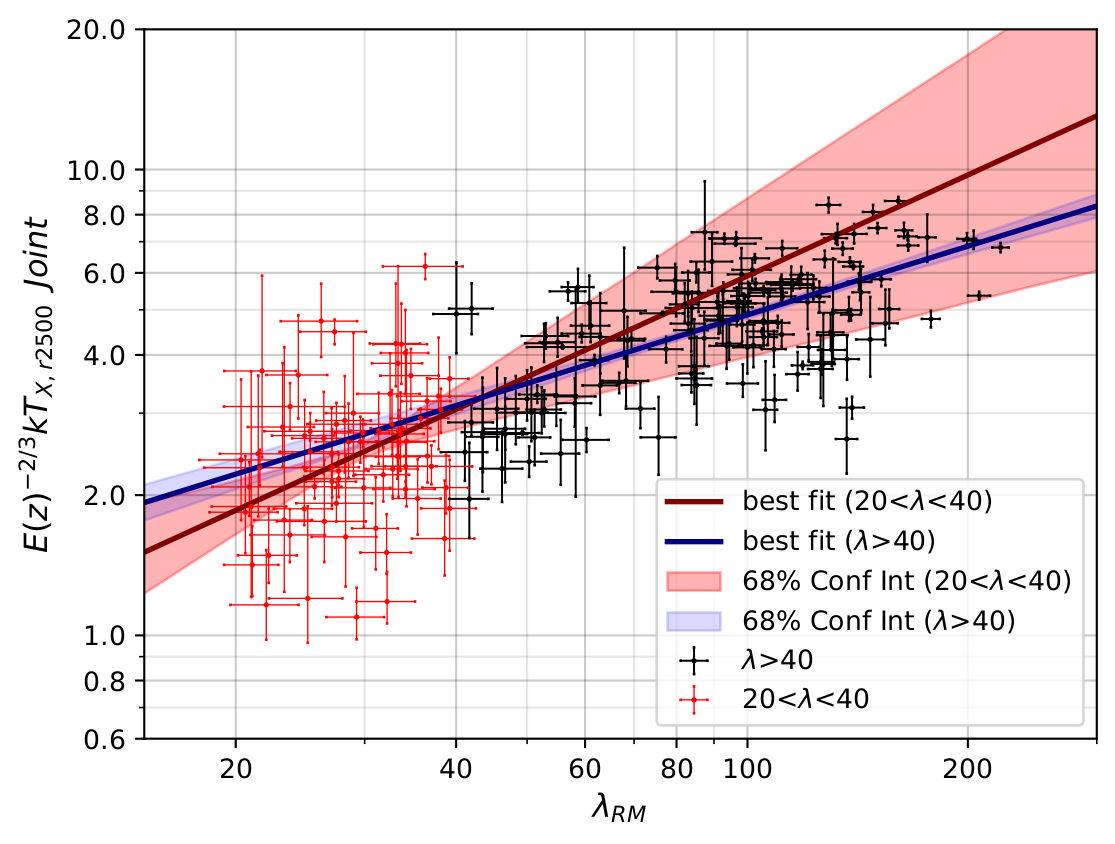} 
    \\[\smallskipamount]
      \small (a) ($20 < \lambda$ $<$ 40) and ($\lambda$ $>$ 40)
    \end{tabular}
    
    \begin{tabular}{@{}c@{}}
      \includegraphics[width=8cm]{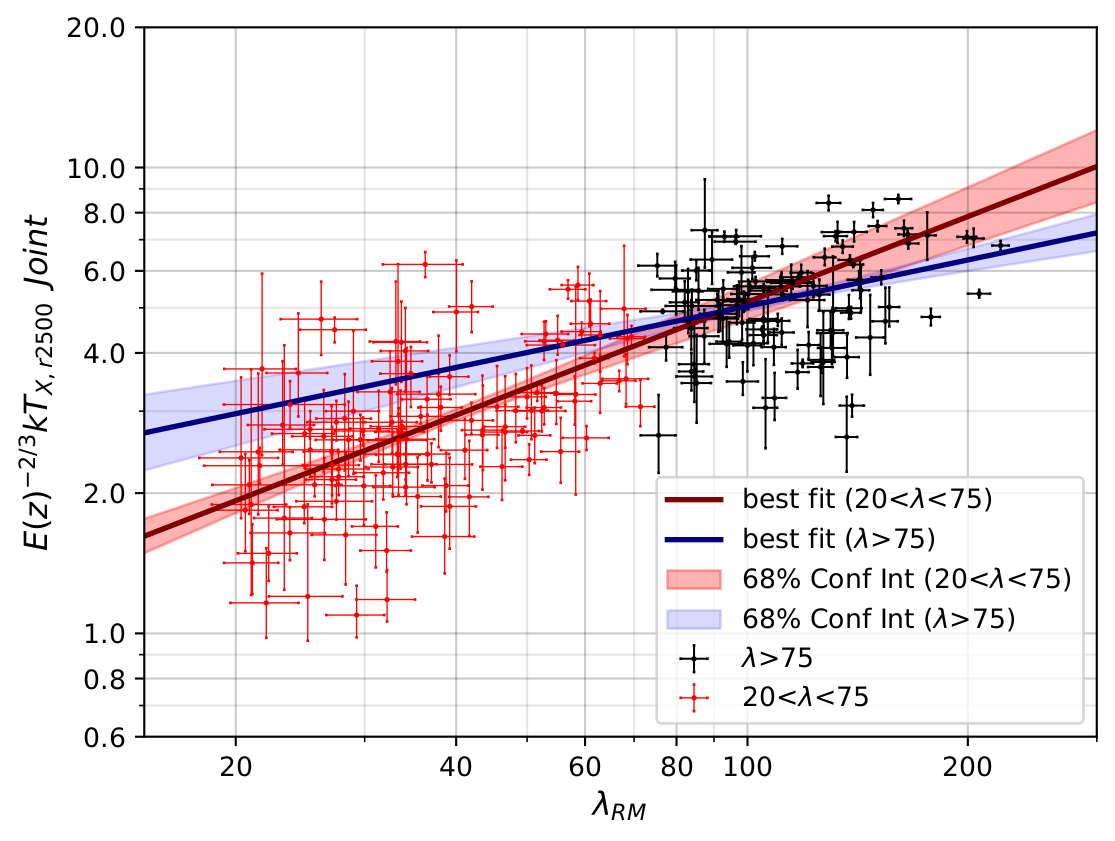} 
    \\[\smallskipamount]
      \small (b) ($20 < \lambda$ $<$ 75) and ($\lambda$ $>$ 75)
    \end{tabular}
    
    \caption{$T_{\rm X}(r_{\rm 2500})-\lambda$ in bins of richness; the top plot (a) shows the fits for the $20 < \lambda$ $<$ 40 and $\lambda$ $>$ 40 bins while the bottom plot (b) shows the comparison of $20 < \lambda$ $<$ 75 and $\lambda$ $>$ 75 clusters. In each case, the red data points and line show clusters in the lower-richness bin and the corresponding best-fit model. Black data points and the blue line show clusters in the higher-richness bin and the best-fit model.}  \label{fig:richnessranges}
\end{figure}

Next we look at the scaling relations in bins of richness.  Here we focus on the $T_{\rm X}-\lambda$ relation only; a richness cut in the $L_{\rm X}-\lambda$ relation results in selecting primarily detected clusters at high richness and primarily undetected clusters at lower richness. In the presented relations, we tried two different sets of richness bins, first cutting at $\lambda$ greater than and less than 40 to separate out particularly low-richness clusters and second cutting at $\lambda$ of 75 giving bins with relatively equal numbers of clusters. The results are presented in Table \ref{table:richness} and shown in Figure~\ref{fig:richnessranges}. 

The large uncertainties on the slope and intercept for the lowest-richness bin $20 < \lambda < 40$ preclude us from finding differences from the higher-richness clusters; however, we do find a marginally higher scatter, at about the $2\sigma$ level, for these low-richness clusters.  These trends are borne out visually in Figure~\ref{fig:richnessranges} (a) where there is no indication of a break in the slope, but the low-richness points do appear to have higher scatter compared to the higher-richness clusters.
For the higher-richness cut of 75, we do find a mildly steeper slope for the lower-richness $20 < \lambda < 75$ clusters compared to the higher-richness clusters, though in this case the scatters are similar.  While the steeper lower-richness slope mirrors the trend for lower-redshift clusters, we note that the median redshifts of all of our richness bins are similar ($z=0.40$) as are the redshift ranges.

\begin{table}
\begin{center}
\begin{tabular}{l c c c c }
\hline
Relation & $\alpha$ & $\beta$ & $\sigma$  & $N$ \\ \hline
$T_{\rm X}$-$\lambda$ (20 $<$ $\lambda$ $<$ 40) & $0.72 \pm 0.32$ & $1.52 \pm 0.28$ & $0.28^{+0.04}_{-0.03}$ & $73$\\
$T_{\rm X}$-$\lambda$ ($\lambda$ $>$ 40) & $0.49 \pm 0.05$ & $1.41 \pm 0.02$ & $0.21 \pm 0.02$ & $152$\\
$T_{\rm X}$-$\lambda$ ($20 < \lambda$ $<$ 75) & $0.61 \pm 0.09$ & $1.42 \pm 0.06$ & $0.25 \pm 0.02$ & $116$\\ 
$T_{\rm X}$-$\lambda$ ($\lambda$ $>$ 75) & $0.33 \pm 0.09$ & $1.50 \pm 0.05$ & $0.21 \pm 0.02$ & $109$\\\hline
\end{tabular}
\caption{$T_{\rm X} (r_{2500})-\lambda$ relation in richness bins; all fits use the joint Chandra and XMM sample.
}
\label{table:richness}
\end{center}
\end{table}

\subsubsection{Serendipitous vs Targeted Clusters}

\begin{figure}
    \centering
    \begin{tabular}{@{}c@{}}
      \includegraphics[width=8cm]{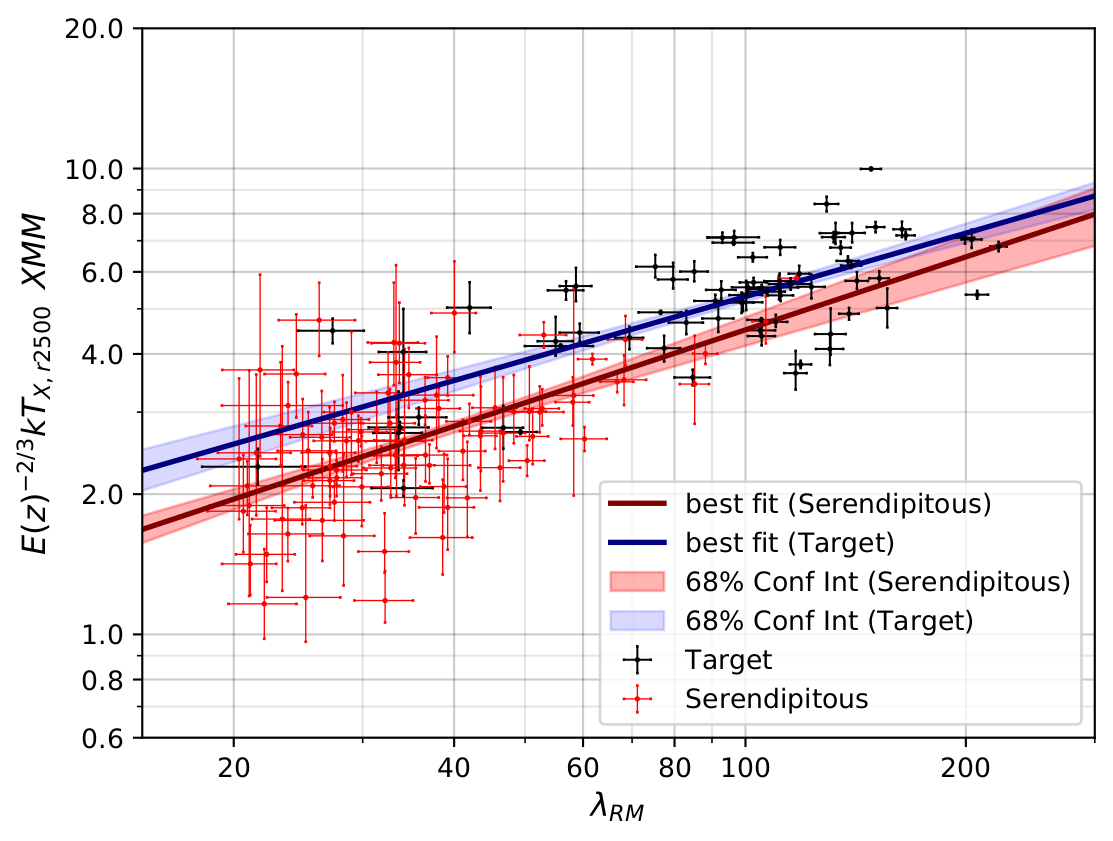}
    \end{tabular}
    \caption{$T_{\rm X}(r_{\rm 2500})-\lambda$ relation for XMM detected clusters only divided according to whether the cluster was the target of the XMM observation or if it was serendipitously detected. Red data points and line show serendipitous clusters and their best-fit scaling relation with the 1 and 2$\sigma$ scatter in yellow. Black data points and the blue line show targeted clusters and the best-fit model with green shaded regions showing the 1 and 2$\sigma$ scatter.}\label{fig:xmmserendipitous}
\end{figure}

In order to explore potential selection effects, we separated the clusters into those that were the target of the observation and those that were detected ``serendipitously". In our archival samples, targeted clusters will be biased toward clusters previously know to be X-ray emitting.  Serendipitous clusters represent a random selection, though we expect the detected serendipitous clusters to be biased toward those more luminous for their richness. As 
the Chandra sample had relatively few serendipitous clusters with temperature measurements, we used only the XMM sample in this test. The results for the $T_{\rm X} (r_{2500})-\lambda$ relation are presented in Table \ref{table:st70} and Figure~\ref{fig:xmmserendipitous}.

The scaling relations for the serendipitous and targeted clusters are consistent with the exception that the target clusters have a higher normalization, making them hotter for their richnesses.  However, the targeted clusters are also on average richer than the serendipitous sample with median richnesses of 105 and 33 for the target and serendipitous samples, respectively. This trend of higher normalization for targeted clusters was also seen in \citet{Giles}.



\begin{table}
\begin{center}
\begin{tabular}{l c c c c }
\hline
Relation & $\alpha$ & $\beta$ &  $\sigma$  & $N$ \\ \hline
Serendipitous & $0.52 \pm 0.07$ & $ 1.32 \pm 0.05$ & $0.18 \pm 0.03$ & $89$\\
Targeted & $0.45 \pm 0.05$ &  $ 1.51 \pm 0.03$ & $0.21 \pm 0.02$ & $71$ \\ \hline
\end{tabular}
\caption{$T_{\rm X} (r_{2500})-\lambda$ relation for the XMM sample only separated based on whether the cluster was serendipitous or targeted.}
\label{table:st70}
\end{center}
\end{table}

\section{Discussion}

Our results confirm the generally good performance of the redMaPPer algorithm.  Our centering results indicate that less than 20\% of redMaPPer clusters are miscentered, and our visual examination revealed that some of the miscentered clusters ($\sim12$\%) were due to gaps in the DES Y3 coverage which will be filled in the final Y6 data set.

We also find a low X-ray temperature-richness scatter of $0.22 \pm 0.01$.  Following the procedure in \citet{farahi19} and using the temperature-mass relation and its scatter from \citet{mantz16} and the richness-temperature correlation coefficient from \citet{farahi19b}, we find a scatter in mass given richness of

\begin{center}
$\sigma_{\ln M|\lambda} = 0.19 \pm 0.02_{\mathrm{stat}} \pm 0.09_{\mathrm{sys}}$
\end{center}

\noindent with the statistical error arising from uncertainty in the $T_X - \lambda$ scatter from our fits and the systematic error arising from uncertainties in the mass-temperature scatter and the richness-temperature correlation.  This mass-richness scatter is consistent with the results of \citet{farahi19}, but our confidence interval favors lower scatter values.
    
However, the DES Y1 cluster cosmology results indicate possible unmodelled systematics in the cluster selection, particularly at low richness \citep{deskp}.  Miscentering, mass-richness scatter, or contamination which grow as richness decreases could potentially contribute to the results seen in DES Y1.  We do see mildly larger scatter in the $T_X-\lambda$ relation in our lowest-richness bin, but not a dramatic increase.  This conclusion is of course limited by the incompleteness of the low-richness sample.  We find no indication of an increase in miscetering at lower richnesses, and in fact the miscentering fraction and distribution are extremely similar in our richness bins.

Looking at the undetected, serendipitous clusters, we find one significantly underluminous cluster; this $\lambda=30$ cluster at $z=0.48$ lies in the same field, but well separated from a high-redshift, Planck cluster with deep Chandra data.  Two other clusters, one detected and one not, have luminosities or limits just outside of three times the $L_X-\lambda$ scatter, but this is also about the number we would expect.
In general, the depth of the X-ray observations is in most cases insufficient to tell whether the undetected clusters are significantly underluminous. \citet{Upsdell23} come to a similar conclusion considering DES redMaPPer clusters in four contiguous XMM survey regions. Looking forward, eROSITA \citep{Merloni12} while not generally deeper will provide complete X-ray coverage with a better understood selection function and much higher sample sizes allowing for studies of redMaPPer selection.  Targeted XMM follow up of underluminous clusters and stacking analyses can also be instructive.

\section{Conclusions}
We investigate the X-ray properties of optically-selected clusters in the first three years of DES data.  Specifically, we analyze 676 redMaPPer-selected DES clusters which fall within archival Chandra and/or XMM observations, of which 239 are detected with SNR$>9$ after flag cuts, in order to probe miscentering, richness scatter, and other aspects of redMaPPer selection.  Our primary results include:

$\bullet$ 10-20\% of redMaPPer clusters are miscentered based on both the X-ray peak to redMaPPer offset distribution and visual inspection.

$\bullet$ The miscentered fraction and typical miscentering distance are consistent in bins of richness and redshift. Given these results and the low miscentering fraction, it is unlikely that miscentering is responsible for the cosmological tension found from DES clusters.

$\bullet$ Of the miscentered clusters, in roughly 40\% the correct central was not a member of any redMaPPer cluster most frequently due to masking because of gaps in the DES coverage or the presence of a nearby bright object.  In two clusters, the presence of an AGN or star formation in the central galaxy caused it to be missed by redMaPPer.  In half of the miscentered cases, the correct central was one of the other four possible centrals identified by redMaPPer.

$\bullet$  Miscentering can lead to underestimates of cluster richness that become significant for large miscentering distances.  Overall, miscentering does not significantly affect the $T_X - \lambda$ relation, but it does lead to a small number of outliers.

$\bullet$ We derive scaling relations between X-ray temperature and luminosity with richness that are generally consistent with previous results.

$\bullet$ We find a $T_{\rm X}(r_{2500}) - \lambda$ scatter of $0.22 \pm 0.01$ for richness calculated at the X-ray peak, and use this to estimate a scatter in mass given richness of $\sigma_{\ln M|\lambda} = 0.19 \pm 0.02_{\mathrm{stat}} \pm 0.09_{\mathrm{sys}}$.

$\bullet$ We find a mildly shallower slope of the $T_{\rm X} - \lambda$ and $L_{\rm X} - \lambda$ relations for higher-redshift clusters ($0.4 < z < 0.65$) than for lower-redshift clusters ($0.2 < z < 0.4$).  We also see a mildly larger scatter for low-richness ($20 < \lambda < 40$) compared to higher-richness ($\lambda > 40$) clusters in the $T_{\rm X} - \lambda$ relation.  The slope and scatter of the $T_{\rm X} - \lambda$ relation are consistent within the errors for serendipitous versus targeted clusters.

$\bullet$ While we see one significantly underluminous cluster given its richness, in general for undetected, serendipitous clusters the X-ray data is not deep enough to probe the purity of low-richness redMaPPer clusters.

Looking forward, LSST and Euclid will provide enormous cluster samples out to much higher redshifts while eROSITA is providing all-sky X-ray coverage.  The utility of optical cluster samples, whether selected by redMaPPer or other cluster finders, will continue to depend on multiwavelength follow up like the work presented here to calibrate cluster selection. X-ray and SZ-selected cluster samples rely on optical confirmation, redshifts, and lensing, making understanding the optical data important for these studies as well.  The large samples present a challenge for follow up efforts, and we are continuing to automate our analysis including many aspects of the visual checking and expanding to the use of eROSITA data.

\section*{Acknowledgements}

This work was supported by the U.S. Department of Energy, Office of Science, Office of
High Energy Physics, under Award Numbers DE-SC0010107 and A00-1465-001.  Funding for the DES Projects has been provided by the U.S. Department of Energy, the U.S. National Science Foundation, the Ministry of Science and Education of Spain, 
the Science and Technology Facilities Council of the United Kingdom, the Higher Education Funding Council for England, the National Center for Supercomputing 
Applications at the University of Illinois at Urbana-Champaign, the Kavli Institute of Cosmological Physics at the University of Chicago, 
the Center for Cosmology and Astro-Particle Physics at the Ohio State University,
the Mitchell Institute for Fundamental Physics and Astronomy at Texas A\&M University, Financiadora de Estudos e Projetos, 
Funda{\c c}{\~a}o Carlos Chagas Filho de Amparo {\`a} Pesquisa do Estado do Rio de Janeiro, Conselho Nacional de Desenvolvimento Cient{\'i}fico e Tecnol{\'o}gico and 
the Minist{\'e}rio da Ci{\^e}ncia, Tecnologia e Inova{\c c}{\~a}o, the Deutsche Forschungsgemeinschaft and the Collaborating Institutions in the Dark Energy Survey. 

The Collaborating Institutions are Argonne National Laboratory, the University of California at Santa Cruz, the University of Cambridge, Centro de Investigaciones Energ{\'e}ticas, 
Medioambientales y Tecnol{\'o}gicas-Madrid, the University of Chicago, University College London, the DES-Brazil Consortium, the University of Edinburgh, 
the Eidgen{\"o}ssische Technische Hochschule (ETH) Z{\"u}rich, 
Fermi National Accelerator Laboratory, the University of Illinois at Urbana-Champaign, the Institut de Ci{\`e}ncies de l'Espai (IEEC/CSIC), 
the Institut de F{\'i}sica d'Altes Energies, Lawrence Berkeley National Laboratory, the Ludwig-Maximilians Universit{\"a}t M{\"u}nchen and the associated Excellence Cluster Universe, 
the University of Michigan, NSF's NOIRLab, the University of Nottingham, The Ohio State University, the University of Pennsylvania, the University of Portsmouth, 
SLAC National Accelerator Laboratory, Stanford University, the University of Sussex, Texas A\&M University, and the OzDES Membership Consortium.

Based in part on observations at Cerro Tololo Inter-American Observatory at NSF's NOIRLab (NOIRLab Prop. ID 2012B-0001; PI: J. Frieman), which is managed by the Association of Universities for Research in Astronomy (AURA) under a cooperative agreement with the National Science Foundation.

The DES data management system is supported by the National Science Foundation under Grant Numbers AST-1138766 and AST-1536171.
The DES participants from Spanish institutions are partially supported by MICINN under grants ESP2017-89838, PGC2018-094773, PGC2018-102021, SEV-2016-0588, SEV-2016-0597, and MDM-2015-0509, some of which include ERDF funds from the European Union. IFAE is partially funded by the CERCA program of the Generalitat de Catalunya.
Research leading to these results has received funding from the European Research
Council under the European Union's Seventh Framework Program (FP7/2007-2013) including ERC grant agreements 240672, 291329, and 306478.
We  acknowledge support from the Brazilian Instituto Nacional de Ci\^encia
e Tecnologia (INCT) do e-Universo (CNPq grant 465376/2014-2).

This manuscript has been authored by Fermi Research Alliance, LLC under Contract No. DE-AC02-07CH11359 with the U.S. Department of Energy, Office of Science, Office of High Energy Physics.

\section*{Data Availability}
The Chandra and XMM catalogs are available in full in machine readable format. The data underlying this paper are available at https://des.ncsa.illinois.edu/releases/dr2/dr2-docs and https://docs.datacentral.org.au/ozdes/overview/ozdes-data-release/. The redMaPPer catalog used is proprietary to the Dark Energy Survey Collaboration, but will be released upon publication of the Y3 cluster cosmology papers.



\bibliographystyle{mnras_2author}
\bibliography{references} 

\section*{Affiliations}
$^{1}$ University of California, Santa Cruz, Santa Cruz, CA 95064, USA\\
$^{2}$ University of California, Davis, 1 Shields Ave, Davis, CA 95616\\
$^{3}$ Santa Cruz Institute for Particle Physics, Santa Cruz, CA 95064, USA\\
$^{4}$ SLAC National Accelerator Laboratory, 2575 Sand Hill Road, Menlo Park, CA 94025\\
$^{5}$ Swinburne University of Technology, John St, Hawthorn VIC 3122, Australia\\
$^{6}$ Department of Physics and Astronomy, Pevensey Building, University of Sussex, Brighton, BN1 9QH, UK\\
$^{7}$ Centre National d'Études Spatiales (CNES), 2, place Maurice-Quentin Paris\\
$^{8}$ Jet Propulsion Laboratory, California Institute of Technology, 4800 Oak Grove Dr., Pasadena, CA 91109, USA\\
$^{9}$ Departments of Statistics and Data Science, University of Texas at Austin, Austin, TX 78757, USA\\
$^{10}$ Kavli Institute for Particle Astrophysics \& Cosmology, P. O. Box 2450, Stanford University, Stanford, CA 94305, USA\\
$^{11}$ SLAC National Accelerator Laboratory, Menlo Park, CA 94025, USA\\
$^{12}$ Los Altos High School\\
$^{13}$ University of California, Berkeley, University Avenue and, Oxford St, Berkeley, CA 94720\\
$^{14}$ University Observatory, Faculty of Physics, Ludwig-Maximilians-Universit\"at, Scheinerstr. 1, 81679 Munich, Germany\\
$^{15}$ Institute of Space Sciences (ICE, CSIC),  Campus UAB, Carrer de Can Magrans, s/n,  08193 Barcelona, Spain\\
$^{16}$ Homestead High School\\
$^{17}$ De Anza College, 21250 Stevens Creek Blvd, Cupertino, CA, USA\\
$^{18}$ Department of Physics and Astronomy, Clemson University, Kinard Lab of Physics, Clemson, SC 29634-0978, US\\
$^{19}$ California State University San Marcos, 333 S Twin Oaks Valley Rd, San Marcos, CA 92096\\
$^{20}$ Dougherty Valley High School\\
$^{21}$ University of New Hampshire, 105 Main St, Durham, NH 03824\\
$^{22}$ Laborat\'orio Interinstitucional de e-Astronomia - LIneA, Rua Gal. Jos\'e Cristino 77, Rio de Janeiro, RJ - 20921-400, Brazil\\
$^{23}$ Fermi National Accelerator Laboratory, P. O. Box 500, Batavia, IL 60510, USA\\
$^{24}$ Department of Physics, University of Michigan, Ann Arbor, MI 48109, USA\\
$^{25}$ Institute of Cosmology and Gravitation, University of Portsmouth, Portsmouth, PO1 3FX, UK\\
$^{26}$ Department of Physics \& Astronomy, University College London, Gower Street, London, WC1E 6BT, UK\\
$^{27}$ Instituto de Astrofisica de Canarias, E-38205 La Laguna, Tenerife, Spain\\
$^{28}$ Universidad de La Laguna, Dpto. Astrofisica, E-38206 La Laguna, Tenerife, Spain\\
$^{29}$ Institut de F\'{\i}sica d'Altes Energies (IFAE), The Barcelona Institute of Science and Technology, Campus UAB, 08193 Bellaterra (Barcelona) Spain\\
$^{30}$ Astrophysics Research Institute, Liverpool John Moores University, Liverpool Science Park, 146 Brownlow Hill, Liverpool L3 5RF, UK\\
$^{31}$ Astronomy Unit, Department of Physics, University of Trieste, via Tiepolo 11, I-34131 Trieste, Italy\\
$^{32}$ INAF-Osservatorio Astronomico di Trieste, via G. B. Tiepolo 11, I-34143 Trieste, Italy\\
$^{33}$ Institute for Fundamental Physics of the Universe, Via Beirut 2, 34014 Trieste, Italy\\
$^{34}$ Hamburger Sternwarte, Universit\"{a}t Hamburg, Gojenbergsweg 112, 21029 Hamburg, Germany\\
$^{35}$ School of Mathematics and Physics, University of Queensland,  Brisbane, QLD 4072, Australia\\
$^{36}$ Institute of Theoretical Astrophysics, University of Oslo. P.O. Box 1029 Blindern, NO-0315 Oslo, Norway\\
$^{37}$ Kavli Institute for Cosmological Physics, University of Chicago, Chicago, IL 60637, USA\\
$^{38}$ Instituto de Fisica Teorica UAM/CSIC, Universidad Autonoma de Madrid, 28049 Madrid, Spain\\
$^{39}$ Center for Astrophysical Surveys, National Center for Supercomputing Applications, 1205 West Clark St., Urbana, IL 61801, USA\\
$^{40}$ Department of Astronomy, University of Illinois at Urbana-Champaign, 1002 W. Green Street, Urbana, IL 61801, USA\\
$^{41}$ School of Mathematics, Statistics, and Computer Science, University of KwaZulu-Natal, Westville Campus, Durban 4041, SA\\
$^{42}$ Center for Cosmology and Astro-Particle Physics, The Ohio State University, Columbus, OH 43210, USA\\
$^{43}$ Department of Physics, The Ohio State University, Columbus, OH 43210, USA\\
$^{44}$ Center for Astrophysics $\vert$ Harvard \& Smithsonian, 60 Garden Street, Cambridge, MA 02138, USA\\
$^{45}$ Australian Astronomical Optics, Macquarie University, North Ryde, NSW 2113, Australia\\
$^{46}$ Lowell Observatory, 1400 Mars Hill Rd, Flagstaff, AZ 86001, USA\\
$^{47}$ George P. and Cynthia Woods Mitchell Institute for Fundamental Physics and Astronomy, and Department of Physics and Astronomy, Texas A\&M University, College Station, TX 77843, USA\\
$^{48}$ George P. and Cynthia Woods Mitchell Institute for Fundamental Physics and Astronomy, and Department of Physics and Astronomy, Texas A\&M University, College Station, TX 77843,  USA\\
$^{49}$ Centro de Investigaciones Energ\'eticas, Medioambientales y Tecnol\'ogicas (CIEMAT), Madrid, Spain\\
$^{50}$ Department of Astronomy, University of Michigan, Ann Arbor, MI 48109, USA\\
$^{51}$ Instituci\'o Catalana de Recerca i Estudis Avan\c{c}ats, E-08010 Barcelona, Spain\\
$^{52}$ Department of Physics, Stanford University, 382 Via Pueblo Mall, Stanford, CA 94305, USA\\
$^{53}$ Department of Physics, Carnegie Mellon University, Pittsburgh, Pennsylvania 15312, USA\\
$^{54}$ Observat\'orio Nacional, Rua Gal. Jos\'e Cristino 77, Rio de Janeiro, RJ - 20921-400, Brazil\\
$^{55}$ Instituto de Fisica Gleb Wataghin, Universidade Estadual de Campinas, 13083-859, Campinas, SP, Brazil\\
$^{56}$ School of Physics and Astronomy, University of Southampton,  Southampton, SO17 1BJ, UK\\
$^{57}$ Computer Science and Mathematics Division, Oak Ridge National Laboratory, Oak Ridge, TN 37831\\
$^{58}$ Instituto de Astrofisica e Ciencias do Espaco, Universidade do Porto, CAUP, Rua das Estrelas, P-4150-762 Porto, Portugal\\
$^{59}$ Departamento de Fisica e Astronomia, Faculdade de Ciências, Universidade do Porto, Rua do Campo Alegre, 687, 4169-007 Porto, Portugal\\
$^{60}$ Lawrence Berkeley National Laboratory, 1 Cyclotron Road, Berkeley, CA 94720, USA




\appendix
\section{Appendix A: Chandra and XMM catalogs}
Table A1 and Table A2 display properties of the galaxy clusters from the XMM and Chandra samples, respectively. The identification for each galaxy is represented in the ``Name'' column, followed by the redMaPPer Mem Match ID, redMaPPer position, richness, redshift, positions of the X-ray peak, the 500 kiloparsec SNR, X-ray observations, and a column to identify serendipitous clusters. Table \ref{table:chandra} has two additional columns to identify the clusters used in the centering analysis and scaling relations.  
\begin{center}
\begin{table*}
\scriptsize{}
\begin{tabular}{l c c c c c c c c c }
\hline
Name & Mem Match ID & RA RM &  Dec RM  & $\lambda$ & $\lambda$ err & z & RA pk & Dec pk & 500kpc SNR \\ \hline
RMJ0254015.5-585710.65	& 1	& 43.5646	& -58.953	& 221.674	& 5.713	& 0.428	& 43.570	& -58.9499	& 9.865	 \\ 
RMJ053255.66-370136.08	& 2	& 83.2319	& -37.0267	& 199.432	& 5.879	& 0.287	& 83.2345	&  -37.0283	& 50.360 \\ 
RMJ230822.21-021131.69	& 3 & 347.0926	& -2.1921 & 163.583	& 4.295	& 0.294	&347.0926 & -2.1922	& 39.792	\\ 
RMJ051637.36-543001.65	& 4	& 79.1557	& -54.5005	& 207.243	& 7.181	& 0.299	& 79.1583	& -54.5145	& 63.349 \\ 
RMJ024524.81-530145.39	& 8	& 41.3534	& -53.0293	& 150.571	& 4.141	& 0.300	& 41.3633	& -53.0314	& 57.125	\\ 
\end{tabular}
\scriptsize{}
\begin{tabular}{l c c c c c c c c c }
\hline
$T_{x,r2500}$  & $T_{x,r2500}\/-$ & $T_{x,r2500}\/+$ & $T_{x,r500}$  & $T_{x,r500}\/-$ & $T_{x,r500}\/+$ &  $L_{x,r2500}$  & $L_{x,r2500}\/-$ & $L_{x,r2500}\/+$ & serend \\ \hline
7.910	& 0.180	& 0.181	& 7.373	& 0.188	& 0.188	& 2.998	& 0.019	& 0.020	& 0\\ 
7.828	& 0.218	& 0.220	& 7.527	& 0.247	&  0.247 & 4.549 & 0.036 & 0.034 & 0\\ 
8.187	& 0.310	& 0.313	& 8.055	& 0.375	& 0.375	& 2.788	& 0.030	& 0.028	& 0\\ 
5.936	& 0.114	& 0.115	& 5.758	& 0.113	& 0.113	& 2.120	& 0.012	& 0.013	& 0 \\ 
8.293	& 0.209	& 0.210	& 8.181	& 0.282	& 0.281	& 4.384	& 0.029	& 0.029	& 0 \\ 
\end{tabular}
\caption{Provided is a sample of XMM galaxy cluster properties. All X-ray temperatures are in units of keV and all X-ray luminosities are in units of $10^{44}$ergs/s. The clusters are identified by their ``Name'' in the first column followed by their ``Mem Match ID'' in ascending order. The ``RA RM'' and ``Dec RM'' columns give the redMaPPer position of the bright central galaxy while ``RA pk'' and ``Dec pk'' give the location of the X-ray peak in each cluster. The richnesses and the 1$\sigma$ richness errors of the clusters are in the ``$\lambda$'' and ``$\lambda$ err'' columns. Following is ``z,'' the redshift column. The 500 kiloparsec SNR is in the ``500kpc SNR'' column. The X-ray temperatures found in an r2500 aperture along with the lower and upper 1$\sigma$ uncertainties are in the ``$T_{x,r2500}$,'' ``$T_{x,r2500}\/-$,'' and ``$T_{x,r2500}\/+$'' columns. Similarly the X-ray temperatures with uncertainty found in an r500 aperture are in the ``$T_{x,r500}$,'' ``$T_{x,r500}\/-$,'' and ``$T_{x,500}\/+$'' columns. The soft band luminosities are in the ``$L_{x,r2500}$'' column, followed by the lower and upper 1$\sigma$ uncertainties in ``$L_{x,r2500}\/-$'' and ``$L_{x,r2500}\/+$,'' respectively. the column titled ``serend'' identifies if a galaxy cluster was found serendipitously. The full catalog of detected XMM clusters is available in full in machine readable format.}
\label{table:xmm}
\end{table*}
\end{center}
\begin{center}
\begin{table*}
\scriptsize{}
\begin{tabular}{l c c c c c c c c c }
\hline
Name & Mem Match ID & RA RM &  Dec RM  & $\lambda$ & $\lambda$ err & z & RA pk & Dec pk & 500kpc SNR \\ \hline
RMJ053255.66-370136.08	 & 2	&83.2319	&-37.0267	&199.432	&5.879	&0.283	&83.2325	&-37.0265	&94.277	 \\ 
RRMJ230822.21-021131.69	& 3	& 347.0926 & -2.1921  & 163.583	& 4.295 & 0.293	& 347.0920	& -2.1912	&136.880 \\ 
RMJ051637.36-543001.65	& 4	& 79.1557	& -54.5005	& 207.243	& 7.181	& 0.299	& 79.1511	& -54.5085	& 58.488 \\ 
RMJ041110.97-481939.64	& 6	& 62.7957	& -48.3277	& 178.045	& 5.029	& 0.413	& 62.8166	& -48.3131	& 78.698 \\ 
RMJ232511.72-411213.33	& 7	& 351.2988	& -41.2037	& 176.001	& 4.716	& 0.358	& 351.2995	& -41.2023	& 39.184\\ 
\end{tabular}
\scriptsize{}
\begin{tabular}{l c c c c c c c c c c c}
\hline
$T_{x,r2500}$  & $T_{x,r2500}\/-$ & $T_{x,r2500}\/+$ & $T_{x,r500}$  & $T_{x,r500}\/-$ & $T_{x,r500}\/+$ &  $L_{x,r2500}$  & $L_{x,r2500}\/-$ & $L_{x,r2500}\/+$ & serend & scaling & center \\ \hline
11.166	&0.638	&0.642	&10.934	&0.858	&0.857	&4.944	& 0.069	&0.068	&0	&1	&1\\ 
nan & nan	& nan	& nan	& nan	& nan &	2.908&	0.021 &	0.021 &	0 &	1 &	1\\ 
10.675	& 0.816	& 0.822	& 14.845	& 1.184	& 2.39	& 3.052	& 0.052	& 0.052	& 0	& 1	& 1\\ 
7.107	& 0.312	& 0.318	& nan	& nan	& nan	& 3.555	& 0.061	& 0.061	& 0	& 1	& 1 \\ 
10.74	& 1.318	& 1.398	& 12.806	& 1.89	& 3.419	& 2.702	& 0.083	& 0.084	& 0	& 1	& 1 \\ 
\end{tabular}
\caption{Chandra galaxy cluster and their properties. All of the column names are identical to those found in Table \ref{table:xmm} with an addition of the columns ``scaling'' and ``centering'' at the end to represent which clusters were used in the centering analysis and the scaling relations. The full catalog of detected Chandra clusters is available in full in machine readable format}
\label{table:chandra}
\end{table*}
\end{center}


\bsp	
\label{lastpage}
\end{document}